\documentstyle[aaspp4,flushrt]{article}

\makeatletter

\newlength{\bibhang}
\let\@internalcite\cite
\def\cite{\@ifstar{\citeyear}{\citefull}}
\def\cite{\let\@citeleft(\let\@citeright)%
    \@ifstar{\citeyear}{\citefull}}
\def\citenp{\let\@citeleft\relax\let\@citeright\relax
    \@ifstar{\citeyear}{\citefull}}
\def\citefull{\def\astroncite##1##2{##1~##2}\@internalcite}
\def\citeyear{\def\astroncite##1##2{##2}\@internalcite}

\def\@citex[#1]#2{\if@filesw\immediate\write\@auxout{\string\citation{#2}}\fi
  \def\@citea{}\@cite{\@for\@citeb:=#2\do
    {\@citea\def\@citea{; }\@ifundefined
       {b@\@citeb}{{\bf ?}\@warning
       {Citation `\@citeb' on page \thepage \space undefined}}
{\csname b@\@citeb\endcsname}}}{#1}}

\def\@cite#1#2{\@citeleft#1\if@tempswa , #2\fi\@citeright}
\def\@biblabel#1{}

\makeatother

\begin{document}
\title{Radiation of Neutron Stars Produced by Superfluid Core}
\author{Anatoly A. Svidzinsky}
\begin{center}
Bartol Research Institute, University of Delaware, Newark, DE 19716, USA \\
asvid@bartol.udel.edu
\end{center}

\begin{abstract}
We find a new mechanism of neutron star radiation wherein radiation is
produced by the stellar interior. The main finding is that neutron star
interior is transparent for collisionless electron sound, the same way as it
is transparent for neutrinos. In the presence of magnetic field the electron
sound is coupled with electromagnetic radiation, such collective excitation
is known as a fast magnetosonic wave. At high densities the wave reduces to
the zero sound in electron liquid, while near the stellar surface it is
similar to electromagnetic wave in a medium. We find that zero sound is
generated by superfluid vortices in the stellar core. Thermally excited
helical vortex waves produce fast magnetosonic waves in the stellar crust
which propagate toward the surface and transform into outgoing
electromagnetic radiation. The magnetosonic waves are partially absorbed in
a thin layer below the surface. The absorption is highly anisotropic, it is
smaller for waves propagating closer to the
magnetic field direction. As a result, the vortex radiation is pulsed with
the period of star rotation. The vortex radiation has the spectral index $
\alpha \approx -0.45$ and can explain nonthermal radiation of middle-aged
pulsars observed in the infrared, optical and hard X-ray bands. The
radiation is produced in the star interior, rather then in magnetosphere,
which allows direct determination of the core temperature. Comparing the
theory with available spectra observations we find that the core temperature
of the Vela pulsar is $T\approx 8\times 10^8$K, while the core temperature
of PSR B0656+14 and Geminga exceeds $2\times 10^8$K. This is the first
measurement of the temperature of a neutron star core. The temperature
estimate rules out equation of states incorporating Bose condensations of
pions or kaons and quark matter in these objects. The estimate also allows
us to determine the critical temperature of triplet neutron superfluidity in
the Vela core $T_c=(7.5\pm 1.5)\times 10^9$K which agrees well with
recent data on behavior of nucleon interactions at high
energies. We also find that in the middle aged neutron stars the vortex
radiation, rather then thermal conductivity, is the main mechanism of heat
transfer from the stellar core to the surface. The core radiation opens a
possibility to study composition of neutron star crust by detection
absorption lines corresponding to the low energy excitations of crust
nuclei. Bottom layers of the crust may contain exotic nuclei with the mass
number up to 600 and the core radiation creates a perspective to study their
properties. In principle, zero sound can also be emitted by other
mechanisms, rather than vortices. In this case the spectrum of stellar
radiation would contain features corresponding to such processes. As a
result, zero sound opens a perspective of direct spectroscopic study of
superdense matter in the neutron star interiors.
\end{abstract}

\section{Introduction}

Properties of matter at densities much larger than nuclear are poorly known
and constitute a challenging problem of modern science. They have broad
implications of great importance for cosmology, the early universe, its
evolution, for compact stars and for laboratory physics of high-energy
nuclear collisions. Present searches of matter properties at high density
take place in several arenas. One of them is investigation of neutron stars.

According to theory of neutron stars (NSs) \cite{Yako99a} a NS can be
subdivided into the atmosphere and four internal regions: the outer crust,
the inner crust, the outer core and the inner core as shown in Fig.~1. The
atmosphere is a thin plasma layer, where thermal electromagnetic radiation
is formed. The depth of the atmosphere varies from some ten centimeters in a
hot NS down to some millimeters in a cold one. The outer crust extends from
the bottom of the atmosphere to the layer of density $\approx 4.3\times
10^{11}{\rm g}$/${\rm cm}^3$ and has a depth of several hundred meters. Ions
and electrons compose its matter. The depth of the inner crust may be as
large as several kilometers. The inner crust consists of electrons, free
neutrons and neutron-rich atomic nuclei. Neutrons in the inner crust are in
a superfluid state. The outer core occupies the density range $0.86-2\psi _0$%
, where $\psi _0=2.8\times 10^{14}{\rm g}$/${\rm cm}^3$ is the saturation
nuclear matter density, and can be several kilometers in depth. It consists
of neutrons $n$ with some (several per cent by particle number) admixture of
protons $p$ and electrons $e$. Almost all theories of NSs predict the
appearance of neutron superfluidity and proton superconductivity in the
outer NS core. In a low-mass NS, the outer core extends to the stellar
center.

More dense NSs also possess an inner core (see Fig.~1). Its radius may reach
several kilometers and central density may be as high as $10-15\psi _0$. The
composition and equation of state of the inner core are poorly known and it
constitutes the main NS ``mystery''. Several hypotheses are discussed in the
literature. One of them is appearance of $\Sigma $- and $\Lambda $-hyperons
in the inner core. The second hypothesis assumes the appearance of pion
condensation. The third hypothesis predicts a phase transition to strange
quark matter composed of almost free $u$, $d$ and $s$ quarks with a small
admixture of electrons. Several authors also considered the hypothesis of
kaon condensation in the dense matter. The radius of the NS usually
decreases as the stellar mass increases (NS becomes more compact). The mass
reaches its maximum at some density, which corresponds to the most compact
stable stellar configuration. Typically the stellar mass ranges between $%
0.5-3M_{\odot }$ for different equations of state ($M_{\odot }$ is the solar
mass).

Despite of 30 years enormous theoretical and observational efforts, the
properties of matter in neutron star cores remain poorly known. The main
reason is that detected thermal radiation from the star surface carries
practically no information about the stellar interior. Some indirect
methods, such as estimates of neutron star masses, radii and cooling rates
for different equation of states, have been applied to find constrains on
the matter behavior. However, non of them is reliable enough because the
answer strongly depends on the model parameters which are also poorly known.

In this paper we discuss a new direct way to determine properties of the
superdense matter. We found that neutron star interior is transparent for
electron zero sound, the same way as it is transparent for neutrinos. Zero
sound is a collective density (or spin-density) wave in Fermi liquids that
was predicted by Landau in 1957 and later observed in $^3$He and electron
liquid in metals. The wave frequency $\omega $ is greater then the frequency
of interparticle collisions which makes the wave different from usual
hydrodynamic sound. In neutron stars the zero sound can propagate in
electron fluid which fills the stellar crust and core, the collisionless
(zero sound) regime is reached at $\omega /2\pi >10^{12}$Hz. The large
density of electrons in NS interior results in substantial reduction of the
interaction energy between fermions as compared to their kinetic (Fermi)
energy, the electron gas becomes more ideal. In such weakly interacting
Fermi system the attenuation of zero sound is exponentially small (see Eq. (%
\ref{at8}) in Appendix A). As a result, zero sound propagates across the
star in a ballistic regime, which is similar to neutrino. However, this is
not the case for single-particle excitations in electron liquid, their mean
free path usually does not exceed $0.01-1{\rm cm}$. The single-particle
excitations are in a local thermodynamic equilibrium with the stellar
matter, while collective zero sound modes do not. At the same time,
negligible attenuation of the zero sound creates an obstacle to excite such
waves inside the stellar medium. Some special processes are required for
their generation. We found that zero sound is generated at the core-crust
interface by helical motion of superfluid vortices. Vortex modes themselves
are excited thermally via single-particle channel, they are in thermodynamic
equilibrium with the surrounding medium.

In the presence of magnetic field the electron sound is coupled with
electromagnetic radiation, such excitation is known as a fast magnetosonic
wave. At large densities the effect of magnetic field is negligible and the
wave behaves as a sound wave, however near the stellar surface (small matter
density) it reduces to electromagnetic wave which leave the star and
propagates into surrounding space. Such property results in a new mechanism
of star radiation produced at the stellar interior. Contrary to neutrino,
electromagnetic radiation is easy to detect which opens a perspective of
direct spectroscopic study of superdense matter. Thermally excited helical
vortex waves generate fast magnetosonic waves in the stellar crust which
propagate toward the surface and transform into outgoing electromagnetic
radiation. We found that such mechanism results in radiation with non
Planck's spectrum (Eqs. (\ref{mss}), (\ref{m104})) and theoretical
predictions agree well with nonthermal radiation of some middle-aged pulsars
observed in the infrared, optical and hard $X$-ray bands (Figs. 5, 6). The
important result of our theory is that detection of vortex radiation in the $%
X-$ray band allows direct determination of the core temperature. From
available spectral data we found that the core temperature of the Vela
pulsar is $T\approx 8\times 10^8$K, while the core temperature of PSR
B0656+14 and Geminga exceeds $2\times 10^8$K. This is the first measurement
of the temperature of a NS core. The cooling rate and temperature of the
central part of a NS substantially depends on the properties of dense
matter. Our temperature estimate excludes exotic equation of states
incorporating Bose condensations of pions or kaons and quark matter in these
objects. Such conclusion agrees with recent measurement of gravitational
redshift of a neutron star \cite{Cott02}. Moreover, the temperature estimate
allows us to determine the temperature of superfluid transition of neutrons
in the Vela core $T_c=(7.5\pm 1.5)\times 10^9$K. This value agrees well with
theoretical predictions that are based on new nucleon-nucleon interaction
potentials which fit the recent world data for $pp$ and $np$ scattering up
to $500{\rm Mev}$ \cite{Bald98,Arnd97}. Examples of such potentials are
Nijmegen I and II. For the matter density $2\psi _0$ (which is a central
region density of a canonical NS without exotic inner core) these nucleon
potentials predict $T_c\approx 7.5\times 10^9$K for neutron superfluidity
with $^3P_2$ (triplet) pairing \cite{Bald98}. One should mention that in
order to find the transition temperature at densities up to $10\psi _0$ more
accurate potentials which fit the measured nucleon-nucleon phase shifts up
to about $1{\rm Gev}$ are required. A systematic study of pulsar radiation
in the $X-$ray band can locate objects with fast cooling core which could be
candidates for NSs with exotic states of matter.

The plan of the paper is the following. In section 2 we discuss neutron
vortices in the superfluid core and generation of magnetosonic waves by
vortices at the crust-core interface. In sections 3, 4 we investigate
propagation of magnetosonic waves across the stellar crust and their
transformation into electromagnetic radiation at the surface. In section 5
we consider spectrum of electromagnetic radiation produced by vortices and
analyze the applicability of our results. In section 6 we compare the theory
with radiation spectrum of some middle-aged pulsars observed in infrared$-X$%
-ray bands and discuss properties of superdense matter based on the observed
spectra. Section 7 is devoted to a problem of heat transport across the
stellar crust. In section 8 we draw our conclusions and discuss perspectives
in the field. In Appendices A, B, C we consider zero sound in an electron
liquid, derive equations of collisionless magnetic hydrodynamics for
electron motion in the stellar crust and investigate propagation of
electromagnetic waves across the atmosphere.

\section{Vortex radiation in rotating neutron stars}

If a NS rotates with an angular velocity ${\bf \Omega =}\Omega \hat z$ a
periodic vortex lattice forms in a superfluid neutron phase of the stellar
core. Typical value of the lower critical angular velocity of the vortex
formation is $\Omega _{c1}/2\pi \sim 10^{-14}s^{-1}$ \cite{Sedr91}. The
vortices are parallel to the axis of rotation apart from a tiny region ($%
\leq 10^{-3}{\rm cm}$) near the condensate surface, where they are curved
and cross the boundary in the perpendicular direction. The distance between
vortices in a triangular lattice is $b=(4\pi \hbar /\sqrt{3}M\Omega )^{1/2}$%
, where $M$ is the mass of the condensate particles. E.g., for a neutron
superfluid ($M=2m_n$, where $m_n$ is the neutron mass) and $\Omega /2\pi =10%
{\rm s}^{-1}$ the lattice period $b=6\times 10^{-3}{\rm cm}$. For NSs the
characteristic size of the vortex core is $\xi =10^{-12}{\rm cm}$, so that $%
b\sim 10^{10}\xi $ and the total number of vortices in the superfluid $%
N=MR^2\Omega /\hbar \approx 10^{17}$, where $R\approx 10{\rm km}$ is the
radius of the superfluid condensate \cite{Sedr91}.

First, let us consider a single straight vortex line with length $L$ located
in a spherical one component superfluid condensate. We assume the condensate
density is uniform inside the sphere. One should note that this situation
differs from real NS core which consists of a mixture of neutrons, protons
and electrons. However, main properties of vortices can be understood within
the simplified model. Normal modes of the vortex are helical waves \cite
{Lifs80}
\begin{equation}
\label{q1}r=A\exp (i\omega t+ik_{\parallel }z),\quad \omega =\frac \hbar {2M}%
\,k_{\parallel }^2\ln (1/k_{\parallel }\xi ),
\end{equation}
where $A$ is the amplitude and $k_{\parallel }$ is the wave number. Eqs. (%
\ref{q1}) are valid for $k_{\parallel }\xi \ll 1$, or $\omega \ll \omega
_c=\hbar /2M\xi ^2\approx 10^{20}$Hz and describe rotation of a helix with
the angular frequency $\omega $.

To find the amplitude of the helical waves $A$ one can use the expression
for the energy of the system in terms of normal vortex modes \cite{Svid00a}.
If a superfluid contains a vortex line of length $L$ it changes the kinetic
energy of the fluid by
\begin{equation}
\label{q11}E_k=\int \frac{MnV^2}2d^3r=\pi Ln\frac{\hbar ^2}M\ln (b/\xi ),
\end{equation}
where $n$ is the superfluid particle density. Change in the kinetic energy (%
\ref{q11}) gives the main contribution to the change in the fluid energy.
One should mention that the superfluid density is suppressed near the vortex
core. However, the normal component fills the vortex core so that the total
fluid density remains approximately constant \cite{Nyga02}. This occurs
because any variation in the total density in a Fermi liquid (under NS
conditions) produces substantial change of the Fermi energy and increases
the energy of the system. The same situation takes place during the vortex
motion and, as a result, the helical vortex waves are not accompanied by
oscillation in the total density. In the wave the superfluid and normal
components oscillate in the opposite phase so that the total fluid density
remains approximately constant.

Taking into account that excitation of the helical wave (\ref{q1}) changes
the vortex length by the value $LA^2k_{\parallel }^2/2$, one can obtain the
energy of the system in terms of normal mode amplitudes \cite{Svid00a}

\begin{equation}
\label{q9}E=E_0+\pi nL\sum_mA_m^2\hbar \omega _m.
\end{equation}
In the NS core the helical vortex waves are excited thermally by particle
scattering on the vortex cores. Vortex excitations are in a local thermal
equilibrium with the stellar matter. Then at temperature $T$ the energy of
the wave with the frequency $\omega $ is equal to $\hbar \omega $ times the
mean number $1/[\exp (\hbar \omega /k_BT)-1]$ of excitations in that state.
Therefore, the vortex wave amplitude is given by (cf. \cite{Bare96})
\begin{equation}
\label{q10}A^2=\frac 1{\pi nL[\exp (\hbar \omega /k_BT)-1]},
\end{equation}
where $k_B$ is Boltzmann's constant. For typical parameters of NSs $A\ll \xi
$. Indeed, if we take $n=10^{39}{\rm cm}^{-3}$, $L=10{\rm km}$, $\omega
/2\pi =10^{15}$Hz, $T=10^8K$ and $\xi =10^{-12}{\rm cm}$, we obtain for the
amplitude of the thermally excited helical wave $A\approx 10^{-9}\xi $.

Apart from helical waves there are excitations of the vortex lattice in a
plane perpendicular to the vortex lines which correspond to relative
displacements of vortices with respect to each other (the so called
Tkachenko's modes). These excitations are analogous to a vortex precession
in trapped Bose-Einstein condensates where a vortex line moves along a
trajectory with constant trapping potential \cite{Svid00b}. However,
frequencies of the vortex oscillation due to excitation of such modes are
small. Indeed, the frequency of Tkachenko's modes is given by $\omega _T=%
\sqrt{\hbar \Omega /4M}k_{\perp }$, where $k_{\perp }$ is the wavenumber in
the plane perpendicular to vortices \cite{Tkac65}. For maximum frequency $%
k_{\perp }\sim 1/b$, where $b$ is the lattice spacing, and we obtain $\omega
_{T\max }\sim \Omega $, which is much less than frequencies of radiation we
are interested in.

\subsection{Energy transport by helical vortex waves in superfluids}

Let us consider a superfluid with a vortex line located along the $z$ axis
and introduce perturbation in the particle density $n^{\prime }$ and the
velocity potential $\chi ^{\prime }$: $n=n_0+n^{\prime }$, $\chi =\chi
_0+\chi ^{\prime }$, where $n_0=n_0({\bf r})$ is the superfluid density for
a straight vortex line and $\chi _0=\hbar \phi /M$. At distances $r\lesssim
\xi $ (we use cylindrical coordinates $r,\phi ,z$ measured from the center
of the vortex line) the helical vortex motion (\ref{q1}) can be described in
terms of $n^{\prime }$ and $\chi ^{\prime }$ as
\begin{equation}
\label{e1}n^{\prime }=-A\exp \left[ i(\omega t+k_{\parallel }z-\phi )\right]
\frac{\partial n_0}{\partial r},
\end{equation}
\begin{equation}
\label{e2}\chi ^{\prime }=iA\frac \hbar {Mr}\exp \left[ i(\omega
t+k_{\parallel }z-\phi )\right] .
\end{equation}
In the bulk the function $\chi ^{\prime }$ (as well as $n^{\prime }$) should
satisfy a linearized equation of superfluid hydrodynamics, which is, in
fact, the wave equation for sound propagation
\begin{equation}
\label{e3}\frac{\partial ^2\chi ^{\prime }}{\partial t^2}-c_s^2\nabla ^2\chi
^{\prime }=0,
\end{equation}
where $c_s$ is the speed of sound in superfluid. In the limit $\omega \ll
\omega _c$ one can omit the first term in Eq. (\ref{e3}). Then solution of
this equation which satisfies the boundary condition (\ref{e2}) has the form
\begin{equation}
\label{e4}\chi ^{\prime }=iA\frac{\hbar k_{\parallel }}MK_1(k_{\parallel
}r)\exp \left[ i(\omega t+k_{\parallel }z-\phi )\right] ,
\end{equation}
where $K_1$ is the modified Bessel function. The solution exponentially
decreases at $r\gg 1/k_{\parallel }$: $K_1(k_{\parallel }r)\approx (\pi
/2k_{\parallel }r)^{1/2}\exp (-k_{\parallel }r)$. We obtain that helical
vortex waves produce perturbation in the superfluid velocity ${\bf V}=\nabla
\chi $ which is localized near the vortex core. Such perturbation results in
the energy flow along the vortex line so that the energy flux is localized
near the vortex.

The energy flux can be estimated as
\begin{equation}
\label{e5}{\bf Q}=\int \frac 12\rho V^{\prime 2}\frac{\partial \omega }{%
\partial {\bf k}}dS,
\end{equation}
where $\rho =Mn$ is the superfluid density, $V^{\prime }$ is the
perturbation in the superfluid velocity due to helical vortex motion and $%
\partial \omega /\partial {\bf k=}\hbar \,k_{\parallel }\ln (1/k_{\parallel
}\xi )\hat z/M$ is the group velocity of the helical wave. In Eq. (\ref{e5})
the integral is taken in a plane perpendicular to the vortex line. Then, we
note that $L\int \frac 12\rho V^{\prime 2}dS$ is the total energy of the
helical wave, which at temperature $T$ is given by
\begin{equation}
\label{e6}L\int \frac 12\rho V^{\prime 2}dS=\frac{\hbar \omega }{\exp (\hbar
\omega /k_BT)-1}.
\end{equation}
Taking Eq. (\ref{e6}) into account, we find the following expression for the
energy flux along the vortex:
\begin{equation}
\label{e7}Q=\frac{\hbar ^{3/2}\omega ^{3/2}\sqrt{\ln (\omega _c/\omega )}}{L%
\sqrt{M}\left[ \exp (\hbar \omega /k_BT)-1\right] }.
\end{equation}
The number of vortex modes within the interval $dk_{\parallel }$ is $%
Ldk_{\parallel }/2\pi =L\sqrt{M}d\omega /2\pi \sqrt{\hbar \omega \ln (\omega
_c/\omega )}$ and, hence, the spectral power density of energy flux along a
single vortex line is
\begin{equation}
\label{e8}P(\omega )=\frac{\hbar \omega }{2\pi \left[ \exp (\hbar \omega
/k_BT)-1\right] }.
\end{equation}
This expression is a universal function in the sense that it depends only on
the temperature. At $\hbar \omega <k_BT$ the energy spectrum is independent
of frequency and $P=k_BT/2\pi $.

One should note that the energy flow produced by helical waves substantially
differs from energy radiation in the case of a moving straight vortex line.
A moving straight vortex in a compressible fluid emits sound waves in the
direction perpendicular to the vortex line. A homogeneous two-dimensional
(2D) superfluid is equivalent to (2+1)-dimensional electrodynamics, with
vortices playing the role of charges and sound corresponding to
electromagnetic radiation \cite{Ambe80,Arov97,Ovch98,Fisc99,Lund00}. Thus,
radiation of sound waves from a straight vortex moving on a circular
trajectory is concentrated in a plane perpendicular to the vortex line and
can be estimated using formulas for the cyclotron radiation of an electrical
charge moving along a circular orbit in 2D space.

To understand the difference between energy radiation by a straight vortex
and a helical wave we can solve the problem in general case. Indeed, general
solution of Eq. (\ref{e3}) that satisfies the asymptotic behavior (\ref{e2})
is
\begin{equation}
\label{e9}\chi ^{\prime }=-\frac{i\pi \hbar A}{2M}k_{\perp }Y_1(k_{\perp
}r)\exp \left[ i(\omega t+k_{\parallel }z-\phi )\right] ,
\end{equation}
where $Y_1$ is the Bessel function and $k_{\perp }=\sqrt{\omega
^2/c_s^2-k_{\parallel }^2}$. At large $r$ the solution (\ref{e9}) has the
asymptotic
\begin{equation}
\label{e10}\chi ^{\prime }=-\frac{i\sqrt{\pi }\hbar A\sqrt{k_{\perp }}}{%
\sqrt{2}M\sqrt{r}}\sin \left( k_{\perp }r-\frac{3\pi }4\right) \exp \left[
i(\omega t+k_{\parallel }z-\phi )\right] ,
\end{equation}
\begin{equation}
\label{e11}n^{\prime }=-\frac n{c_s^2}\frac{\partial \chi ^{\prime }}{%
\partial t}=-\frac{\sqrt{\pi }\hbar \omega nA}{\sqrt{2}Mc_s^2\sqrt{r}}\sqrt{%
k_{\perp }}\sin \left( k_{\perp }r-\frac{3\pi }4\right) \exp \left[ i(\omega
t+k_{\parallel }z-\phi )\right] .
\end{equation}
For $\omega >c_sk_{\parallel }$ ($k_{\perp }$ is a real number) the solution
describes radiation of sound waves with a conical wavefront (the cone angle
is $\arctan (k_{\perp }/k_{\parallel })$). The particular case $k_{\parallel
}=0$ corresponds to a circular motion of a straight vortex and sound waves
are radiated in the plane perpendicular to the vortex line. However, in the
case of helical waves the wavelength of sound that would be emitted is much
larger than the wavelength of helical wave and the condition of sound
radiation $\omega >c_sk_{\parallel }$ (which for a helix is equivalent to $%
\omega >4\omega _c/\ln (\omega _c/\omega )$) fails. As a result, the
solution (\ref{e9}) reduces to (\ref{e4}) which describes energy flow along
the vortex.

\subsection{Magnetic field associated with neutron vortices}

Now let us take into account a superconducting proton component in NS core.
The bulk of superconducting protons do not rotate by forming vortices, which
differs from neutron superfluid. Superconducting protons corotate with the
crust and electron fluid by generating a surface current which produces a
small uniform magnetic field $\,$in the interior ${\bf H}=-2m_pc{\bf \Omega }%
/e$, here $m_p$, $e$ is the proton mass and charge \cite{Alpa84}. The vector
potential associated with this field ${\bf A=-}m_pc({\bf \Omega }\times {\bf %
r)/}e$ yields the proton velocity which corresponds to rigid body rotation%
$$
{\bf V}_p=\frac \hbar {2m_p}\left( \nabla \varphi -\frac{2e}{\hbar c}{\bf A}%
\right) ={\bf \Omega }\times {\bf r},
$$
while the phase of the proton order parameter $\varphi $ is constant inside
the superconductor. However, due to interactions between protons and
neutrons, neutron superfluid velocity generates a superfluid current of
protons in the vicinity of neutron vortices (drag effect) and an associated
magnetic field around each neutron vortex line. The magnetic field of a
neutron vortex is confined within a radius of the order of the effective
London length $\Lambda _{*}^2=m_pm_p^{*}c^2/4\pi e^2\rho _p$, where $\rho _p$
is the mass density of superfluid protons and $m_p^{*}$ is the proton
effective mass \cite{Alpa84}. For the interior density $\rho _n=10^{14}{\rm %
g/cm}^3$ and proton concentration $\rho _p/\rho _n=0.05$ the screening
length is $\Lambda _{*}^{}\approx 10^{-11}{\rm cm}$. At $r\geq \xi $ the
magnetic field is given by
\begin{equation}
\label{m0}{\bf H}_v=\frac{\Phi _{*}K_0(r/\Lambda _{*})}{2\pi \Lambda _{*}^2}%
\hat z,
\end{equation}
where $\Phi _{*}=\Phi _0\delta m_p^{*}/m_n$, $\Phi _0=\pi \hbar c/e$ is the
flux quantum, $\delta m_p^{*}=m_p^{*}-m_p$ is the contribution to the proton
effective mass due to the interaction with the neutron medium and $K_0$ is
the modified Bessel function. Sj\"oberg (1976) has shown that for typical
barion densities of a NS core $1-6\times 10^{14}{\rm g/cm}^3$ the ratio $%
m_p^{*}/m_p$ changes approximately from $0.5$ to $0.25$. This result is
important for our theory since it indicates that neutron vortices are
strongly coupled with protons that carry electric charge. As a consequence,
motion of neutron vortices is accompanied by motion of protons and,
therefore, electrons. This creates a possibility of excitation collective
modes in the electron liquid by oscillating neutron vortices. In
superconducting ($npe$) phase the magnetic field lines are parallel to
neutron vortices which is similar to magnetic field of a long solenoid with
the radius $\Lambda _{*}^{}$. In the stellar crust there is no drag effect
and magnetic lines diverge the same way as magnetic lines near the edge of
the solenoid.

One should note that fossil (not associated with neutron vortices) magnetic
field can exist in the stellar core. However, the process of the magnetic
fields generation in NSs and, hence, their structure and localization is
still under discussion. There are arguments that NS magnetic fields may be
confined to the crustal layer only, where it has been generated by
thermomagnetic instability during the first years of NS evolution \cite
{Blan83,Urpi86,Wieb96}. However, there is a possibility that magnetic field
permeates both the stellar crust and the core. Such magnetic field (if it
exists) forms before the transition into superconductivity by convection
which arises in first 10-20 seconds after the NS is born and most probably,
has a very complicated twisted structure \cite{Thom93}. As a rule the
magnetic flux tubes inside a neutron star do not parallel to the axis of
rotation \cite{Rude98}. One should mention, that typical value of the
critical field for vortex formation in superconducting core, $10^{15}${\rm Gs%
}, is much greater than actual magnetic field in NSs $10^{12}${\rm Gs}.
Therefore, the core state with penetrating fossil magnetic field is
metastable. Proton vortices are expelled from superconductor by different
dissipation mechanisms which determine the life time of the metastable
configuration. One of the mechanisms is the following. Since the neutron
vortex and a flux tube strongly interact as they pass through each other,
the moving neutron vortices will push on the proton's flux-tube array,
forcing it to move outside in a spinning-down NS \cite{Saul89}. As neutron
vortex motion moves an entrained flux tube, that tube is ultimately pushed
into the crust-core interface for almost any initial flux-tube
configuration. E.g., during the spinning-down stage when the initial stellar
rotation period of $1${\rm ms} (at the star birth) increases to $100${\rm ms}
(middle aged pulsars) the spacing between neutron vortices increases by a
factor of $10$. As a result all magnetic tubes should be pushed out into the
crust unless originally some of them pass very close (at a distance less
then $0.1$ stellar radii) to the axis of rotation.

In this paper we will not consider possible presence of fossil magnetic
tubes in the stellar core. Such tubes would not modify radiation produced by
neutron vortices, however, they may give an additional contribution into
stellar radiation. The main reason why we do not consider the possible
presence of fossil magnetic tubes is that the observed spectra of middle
aged pulsars can be explained well only by radiation from neutron vortices.
The other argument is that the core temperature we determined from the
observed spectra is very high. The big difference between the core and
surface temperatures suggests strong heat isolation of the stellar core. In
Sec. 7 we show that geometry in which magnetic lines at the crust bottom are
parallel to the core surface (or magnetic field has a closed configuration
within the star) provides sufficient heat isolation and can explain the
observations. Such geometry suggests that fossil magnetic field, if it has
been generated in the core, already expelled from the superconducting core
at the stage of middle aged pulsars. Based on the results of this paper we
can provide another argument against presence of fossil magnetic field in
the stellar core. The point is that our findings show that superfluidity of
neutrons and superconductivity of protons in the NS\ core are very strong.
Such conclusion is also supported by the theory which is based on new
realistic interaction potentials between nucleons \cite{Bald98,Elga96}. The
temperature of superconducting transition for protons near the stellar
center is expected to be no less than $2\times 10^{10}$K \cite{Elga96}. The
modified Urca process cools the stellar core to such temperature during the
time $\tau \approx 1{\rm yr}/T_9^6=0.5${\rm s }\cite{Peth92}. This means
that stellar interior practically instantly after the star formation becomes
superconducting and the convection mechanism cannot generate magnetic field
at the stellar core simply because there is no time for that.

\subsection{Generation of magnetosonic waves at the boundary of
superconducting phase}

At the interface between the stellar crust and the outer core (at the
density $2.4\times 10^{14}{\rm g/cm}^3$) the matter undergoes a first order
transition from a nucleus-electron-neutron ($Aen$) phase to a uniform liquid
of neutrons, protons and electrons ($npe$ phase). Nuclei in the $Aen$ phase
can be spherical and form a Coulomb crystal or have a deformed shape which
results in an exotic distributions of nuclear matter. Transition from the $%
Aen$ phase to the uniform liquid is accompanied by a small (about $1.4\%$)
density jump \cite{Douc00}. In the $npe$ phase the proton liquid is
superconducting \cite{Baym69}, while in the $Aen$ phase protons constitute a
part of nuclei. As a result, the $Aen$ phase is not superconducting,
however, neutron component is still superfluid.

As we have shown before, helical vortex motion does not generate sound waves
in the stellar core because the condition of sound radiation is not
satisfied. However, at the interface between the superconducting ($npe$) and
$Aen$ phases such vortex motion produces magnetosonic waves which propagate
across the stellar crust toward the surface. One should note that
frequencies of radiation we are interested in are larger than effective
frequencies of electron-phonon and electron-electron collisions $\nu _{{\rm %
e,ph}}$, $\nu _{{\rm e,e}}$. According to Eqs. (\ref{d23}), (\ref{d232}) at $%
T=10^8$K, $\rho =10^{14}{\rm g/cm}^3$, $n_e=10^{37}{\rm cm}^{-3}$ for $%
_{26}^{56}$Fe lattice we obtain $\nu _{{\rm e,ph}}=8\times 10^{11}{\rm s}%
^{-1}$, $\nu _{{\rm e,e}}=10^{10}{\rm s}^{-1}$. So, at $\omega /2\pi
>10^{12} ${\rm Hz} the electrons in the stellar crust behave as an
independent system of Fermi particles and move in collisionless regime. In
such regime a zero sound is a possible collective excitation of Fermi
liquid. In details the zero sound in electron liquid is discussed in
Appendix A. In the presence of magnetic field the electron sound is coupled
with electromagnetic radiation. In Appendix B we derive equations of
electron motion in collisionless regime and show that they reduce to usual
equations of magnetic hydrodynamics in which speed of zero sound enters the
equations instead of speed of usual sound and the Alfven velocity is
estimated in terms of the electron density $\rho _e=m_en_e$: $u_A=H/\sqrt{%
4\pi \rho _e}$. One should mention that near the crust bottom the Alfven
velocity is small because of large density. If $H=10^{12}{\rm Gs}$ and $\rho
_e=10^{11}{\rm g/cm}^3$ we obtain $u_A=9\times 10^5{\rm cm/s}$ which is much
less than the speed of zero sound $u_s\approx c $, where $c$ is the speed of
light. In magnetic hydrodynamics there are three possible types of waves:
Alfven, fast and slow magnetosonic. Here we consider radiation of fast
magnetosonic waves which are in the limit $u_s\gg u_A$ reduce to zero sound.
Vortex motion also generates Alfven and slow magnetosonic waves in the
stellar crust. However, the intensity of radiation of those waves is small
(in parameter $u_A^3/u_s^3$) as compared to the intensity of fast
magnetosonic wave. Also Alfven and slow magnetosonic waves are highly damped
and can not reach the stellar surface. Due to these reasons we will not
discuss radiation of Alfven and slow magnetosonic waves.

Let us now consider excitation of electron zero sound by vortices in detail
(see Fig. 2). Helical waves produce oscillation of the neutron vortex and,
therefore, oscillation of protons in the $npe$ phase. In $Aen$ phase there
is no drag effect and no electric current is produced by neutron vortices.
For helical vortex motion the superfluid velocity of neutrons is
$$
{\bf V}_n=\nabla \chi =\frac \hbar {2m_nr}\hat \phi +{\bf V}_n^{\prime }
$$
where ${\bf V}_n^{\prime }=\nabla \chi ^{\prime }$ is the correction to the
velocity due to the helical vortex motion,
\begin{equation}
\label{v0}\qquad \chi ^{\prime }=iA\frac \hbar {2m_nr}\exp \left[ i(\omega
t+k_{\parallel }z-\phi )\right] .
\end{equation}
In the $npe$ phase at $r<\Lambda $, where $\Lambda $ is the screening
length, the induced proton velocity around the neutron vortex is given by
\cite{Alpa84}:

\begin{equation}
\label{v1}{\bf V}_p\approx \frac{\delta m_p^{*}}{m_p^{*}}{\bf V}_n{\bf =}%
\frac{\delta m_p^{*}}{m_p^{*}}\frac \hbar {2m_nr}\hat \phi +{\bf V}%
_p^{\prime }{\bf (}t),
\end{equation}
where ${\bf V}_p^{\prime }{\bf =}\delta m_p^{*}{\bf V}_n^{\prime }{\bf /}%
m_p^{*}$. The last term in (\ref{v1}) is time dependent and produces density
oscillation of protons according to the continuity equation
\begin{equation}
\label{v2}\frac{\partial \rho _p}{\partial t}=-\rho _p{\rm div}{\bf V}%
_p^{\prime }.
\end{equation}
One should note that only $z$ component of ${\bf V}_p^{\prime }$ contributes
to the density oscillation since ${\rm div}{\bf V}_{p\perp }^{\prime }{\bf =}%
0$. The Debye length for electrons, $\Lambda _e=\sqrt{m_ec^2/4\pi e^2n_e}$,
is much less than characteristic scales of the proton density variation. As
a result, free electrons screen the electric field produced by the moving
protons, that is average electron velocity is approximately equal to the
oscillating part of the proton velocity. In $Aen$ phase there is no drag
effect and electron motion is not coupled with neutron vortices.

We will assume that the core-crust interface is transparent for electrons
and they can freely move from one phase into another. The assumption is
reasonable since the electron Fermi energy is much larger than any Coulomb
barrier that might exist at the interface over the interparticle scale. At
the interface the oscillating electron motion generates zero sound waves
which propagate across the stellar crust. To estimate the power of the
radiated sound one can use the following simple analysis. We note that
kinetic energy of ultrarelativistic electrons at the crust bottom is much
larger than their interaction energy (Fermi gas becomes more ideal with
increasing its density). Therefore, the radiated power of zero sound should
be approximately the same as in the limiting case of non interacting
electrons. In this limit, the electrons move in a ballistic regime and the
energy flux across the crust-core interface is given by
\begin{equation}
\label{v4}Q=\int E_ev_FdS,
\end{equation}
where $v_F\approx c$ is the electron Fermi velocity and $E_e$ is the
contribution to the electron energy density due to the helical vortex
motion. The integral in Eq. (\ref{v4}) is the surface integral over the
interface. Eq. (\ref{v4}) takes into account that only electrons that move
along the vortex line are coupled with the helical vortex motion and carry
its energy. The rest electrons can not be excited by the helical wave
because it would mean energy radiation by the helix in a perpendicular plane
which is not possible (see Sec. 2.1).

As we have shown above, see Eq. (\ref{e6}), for a vortex in a single
component superfluid the change in kinetic energy of neutrons determines the
energy of the helical wave. However, this is not the case for a neutron
vortex in the stellar core. Due to the drag effect the helical vortex motion
inevitably produces density oscillation of superconducting protons and
normal electrons (as we have mentioned in Sec. 2.2 the drag is very strong).
As a result, the density oscillation of electrons (which is accompanied by
large change in their Fermi energy) gives the main contribution to the helix
energy in the stellar core and, hence, the energy of the helical wave is
equal to $\int E_eLdS$, where $L$ is the vortex length.

Taking into account Eq. (\ref{v4}), we obtain the following average energy
flux of zero sound waves produced by a helical wave of a single vortex:
\begin{equation}
\label{v5}Q_1=\frac cL\int E_eLdS=\frac cL\frac{\hbar \omega }{[\exp (\hbar
\omega /k_BT)-1]}.
\end{equation}
One should note that the wave length of a sound is much larger than the size
$\Lambda $ of the area from which it is generated. Therefore, the sound wave
has approximately a spherical front (which is similar to diffraction of
light on a small orifice).

The number of vortex modes within the interval $dk_{\parallel }$ is equal to
$Ldk_{\parallel }/2\pi =L\sqrt{2m_n}d\omega /2\pi \sqrt{\hbar \omega \ln
(\omega _c/\omega )}$ and, therefore, the spectral power density of sound
waves produced by a single vortex line is
\begin{equation}
\label{v7}P_1(\omega )=\frac{c\sqrt{\hbar m_n\omega }}{\sqrt{2}\pi \sqrt{\ln
(\omega _c/\omega )}\left[ \exp (\hbar \omega /k_BT)-1\right] }.
\end{equation}
Eq. (\ref{v7}) is a universal expression that actually depends only on the
core temperature $T$ and the neutron mass $m_n$. Multiplying Eq. (\ref{v7})
by the number of neutron vortices in the superfluid $N=2m_nR^2\Omega /\hbar $
and dividing by $2\pi $, we obtain the spectral density of sound waves power
radiated from the oscillating vortex lattice in unit solid angle
\begin{equation}
\label{mss}P_{{\rm v}}(\omega )=\frac{cm_n^{3/2}R^2\Omega \sqrt{\omega }}{%
\sqrt{2}\pi ^2\sqrt{\hbar \ln (\omega _c/\omega )}\left[ \exp (\hbar \omega
/k_BT)-1\right] },
\end{equation}
where $T$ is the temperature of the NS core.

Fast magnetosonic waves generated by vortices at the core-crust interface
propagate across the stellar crust toward the surface. Along the wave
propagation the properties of medium do not change very much over the length
of the wave and the approximation of geometric acoustics is valid. The
energy propagates along the rays and in the absence of absorption the energy
flux remains constant along the direction of the ray, while properties of
the medium slowly vary in space.

\section{Attenuation of fast magnetosonic waves in the stellar crust}

In this section we estimate attenuation of fast magnetosonic waves
propagating across the stellar crust. We assume that crust matter is
everywhere in a crystal phase: $T<T_m$, where $T_m$ is the melting
temperature. For iron crust $T_m=2.5\times 10^7\rho _6^{1/3}$K, where $\rho
_6$ is the matter density in units $10^6{\rm g/cm}^3$ \cite{Hern84}. Near
the stellar surface there is a phase transition between a gaseous atmosphere
and a condensed metallic phase. For iron the critical temperature of the
phase transition can be estimated as $T_{{\rm crit}}\sim
10^{5.5}H_{12}^{2/5} $K, where $H_{12}=H/10^{12}$, which is larger than the
surface temperature ($\sim 10^6$K) for strong enough magnetic field. We also
assume that the atmosphere above the condensed surface has negligible
optical depth which is the case when $T\lesssim T_{{\rm crit}}/3$ \cite
{Lai01}. Under such conditions electromagnetic radiation propagates through
the atmosphere without damping. In Appendix C we study in detail attenuation
of electromagnetic waves in hydrogen atmosphere and show that atmosphere is
transparent for electromagnetic radiation if the density at the atmosphere
bottom $\rho \lesssim 4{\rm g/cm}^3$. The density of the metallic phase at
the stellar surface depends on magnetic field. E.g., for iron crust and $%
H=10^{12}{\rm Gs}$ the matter density at the crust surface is $\rho \approx
10^3{\rm g/cm}^3$ \cite{Thor98}.

In our problem the speed of fast magnetosonic wave $u\,$is larger than the
electron Fermi velocity (see Appendix B) and the speed of usual sound in the
crystal lattice (phonons). Under such conditions the simple absorption
processes (Landau damping) of a fast magnetosonic wave by electrons and
phonons are not allowed due to a requirement of energy and momentum
conservation. In Appendix A we discuss attenuation of zero sound and
demonstrate that it is exponentially small for a weakly interacting Fermi
gas, such as a gas of electrons in the NS crust and core (see Eq. (\ref{at8}%
)). This contribution into attenuation comes from the collision integral in
the kinetic equation. In the presence of magnetic field the electron sound
is coupled with electromagnetic waves which results in another mechanism of
damping. The point is that the fast magnetosonic waves are accompanied by
oscillation of the electric field and in the presence of finite electrical
conductivity $\sigma $ the magnetosonic waves slightly dissipate energy.
This contribution dominates and we discuss it in this section. To find the
wave attenuation one should solve equations of magnetic hydrodynamics for
electron motion with the resistive term (see Eq. (\ref{t5})), they are
derived in Appendix B. As a result we obtain the following dispersion
equation for the magnetosonic branch
\begin{equation}
\label{d8}\omega ^2(\omega ^2-k^2u_s^2)\left( 1-\frac{ic^2}{4\pi \sigma
\omega }\left( k^2-\frac{\omega ^2}{c^2}\right) \right) -(\omega
^2-k_{\parallel }^2u_s^2)\left( k^2-\frac{\omega ^2}{c^2}\right) u_A^2=0,
\end{equation}
where $u_s\approx v_F$ is the speed of zero sound, $u_A=H/\sqrt{4\pi \rho _e}
$ is the Alfven velocity, $\rho _e$ is the electron density and $%
k_{\parallel }$ is the component of the wavevector along the magnetic field $%
{\bf H}$.

In the metallic crust the scattering of electrons on phonons is the dominant
electron scattering mechanism. We will not discuss scattering on impurities
since strong gravitational field produces space separation of nuclei with
different masses during the ``hot'' stage of crust evolution. As a result,
concentration of nuclei with different charges immersed in lattice sites is
expected to be negligible. Effect of magnetic field on electron scattering
is negligible unless the system as a whole occupies only a small number of
Landau levels. Quantitatively this condition means that the number of levels
populated is less than about $10$ \cite{Hern84}. For iron lattice, the
condition is equivalent to
\begin{equation}
\label{d5}\rho _6<0.6H_{12}^{3/2}.
\end{equation}
When condition (\ref{d5}) is satisfied the electron motion becomes
effectively one dimensional. However, in 1D the simple electron-phonon
scattering is not allowed because it is impossible to satisfy equations of
energy and momentum conservation. Therefore, at densities given by Eq. (\ref
{d5}) the electron scattering and magnetosonic attenuation is suppressed by
magnetic field. The wave attenuation can be substantial at larger densities
where many Landau levels are populated and the magnetic field does not
suppress the electron scattering. Let us now focus on this region. In this
region the electron Fermi momentum is given by $p_F\approx \hbar (3\pi
^2n_e)^{1/3}$ and the Alfven velocity $u_A$ is smaller than $u_s\approx v_F$
(which is the case when $\rho _6>0.16H_{12}^2A/Z$, where $Z$ is the nuclear
charge, $A$ is the atomic mass number).

At $u_A\ll u_s$ we obtain from Eq. (\ref{d8}) the following correction to
the frequency of fast magnetosonic wave
\begin{equation}
\label{d91}\omega =ku_s+\frac{i(c^2-u_s^2)^2\omega ^2u_A^2\sin ^2\theta }{%
8\pi \sigma u_s^4c^2},
\end{equation}
where $\theta $ is the angle between the wavevector ${\bf k}$ and the
magnetic field ${\bf H}$, $u_s\approx v_F$ is the speed of the wave. The
wave propagates without damping a distance $l$ which depends on electron
conductivity:
\begin{equation}
\label{d11}l\approx \frac{u_s}{{\rm Im}\omega }=\frac{8\pi \sigma u_s^5c^2}{%
\omega ^2(c^2-u_s^2)^2u_A^2\sin ^2\theta }.
\end{equation}
In the presence of magnetic field the electron conductivity can be estimated
as \cite{Kitt68}
\begin{equation}
\label{d12}{\rm Re}\frac 1{\sigma _{\perp }}={\rm Re}\frac 1{\sigma _0}%
\left( 1+\frac{\omega _{{\rm He}}^2}{\omega ^2+\nu _{{\rm eff}}^2}\right)
,\quad {\rm Re}\frac 1{\sigma _{\parallel }}={\rm Re}\frac 1{\sigma _0},
\end{equation}
where $\sigma _0\approx e^2c^2n_e/E_F\nu _{{\rm eff}}$ is the static
conductivity without magnetic field, $\omega _{He}=eHc/E_F$ is the electron
cyclotron frequency, $E_F=\sqrt{m_e^2c^4+c^2p_F^2}$ is the electron Fermi
energy, $\nu _{{\rm eff}}$ is the effective frequency of electron
collisions, $\sigma _{\perp }$, $\sigma _{\parallel }$ are conductivities
across and along the field ${\bf H}$. In magnetosonic waves the oscillating
electric field is perpendicular to the magnetic field ${\bf H}$. Therefore, $%
\sigma _{\perp }$ should be used to estimate the attenuation.

Also in the considering density region the crust temperature is much less
then the Debye temperature \cite{Kitt68}
\begin{equation}
\label{d22}T_D=\frac{(6\pi ^2n_i)^{1/3}\hbar c_s}{k_B}=9.7\times 10^7\rho
_6^{1/2}\frac{Z^{2/3}}A{\rm K},
\end{equation}
where $c_s$ is the speed of usual sound, $n_i$ is the ion concentration. In
the limit $T\ll T_D$ there are two effective frequencies of electron-phonon
collisions \cite{Lifs79}. One of them $\nu _{1{\rm e,ph}}$ describes energy
relaxation (fast relaxation) and responsible for thermal conductivity in the
absence of magnetic field
\begin{equation}
\label{d23}\nu _{1{\rm e,ph}}\sim \frac{k_BT}{2\pi \hbar }\left( \frac T{T_D}%
\right) ^2=\frac{2\times 10^{12}T_6^3A^2}{\rho _6Z^{4/3}}{\rm s}^{-1}.
\end{equation}
The other one $\nu _{2{\rm e,ph}}$ describes slow relaxation of electrons in
momentum directions and determines electrical conductivity
\begin{equation}
\label{d231}\nu _{2{\rm e,ph}}\sim \frac{k_BT}{2\pi \hbar }\left( \frac T{T_D%
}\right) ^4=\frac{2\times 10^9T_6^5A^4}{\rho _6^2Z^{8/3}}{\rm s}^{-1}.
\end{equation}
Using Eqs. (\ref{d11}), (\ref{d12}), (\ref{d231}) in the limit $\nu _{{\rm %
eff}}\ll \omega \ll \omega _{He}$ we obtain for ultrarelativistic electrons
(that is when $\rho _6>A/1.03Z$) the following expression for the free path
of fast magnetosonic waves
\begin{equation}
\label{d24}l\approx \frac{8\pi c^9n_ee^2}{(c^2-v_F^2)^2u_A^2\omega _{{\rm He}%
}^2E_F\nu _{2{\rm e,ph}}\sin ^2\theta }\approx \frac{1.14\times 10^3\rho
_6^{17/3}Z^{19/3}}{H_{12}^4T_6^5A^{23/3}\sin ^2\theta }{\rm cm}.
\end{equation}

The matter density at the stellar crust $\rho $ as a function of the
distance to the surface $h$ can be obtained from the condition that
hydrostatic pressure $\int \rho gdh$ is equal to the pressure of electrons
(here $g$ is the acceleration of gravity). The pressure of ultrarelativistic
electron gas is given by
\begin{equation}
\label{d20}P=\frac{(3\pi ^2)^{1/3}}4\hbar cn_e^{4/3}=\frac{(3\pi
^2)^{1/3}\hbar c}{4m_p^{4/3}}\rho ^{4/3}\left( Z/A\right) ^{4/3}
\end{equation}
and we estimate
\begin{equation}
\label{d241}h=\frac{(3\pi ^2)^{1/3}\hbar c\rho ^{1/3}}{gm_p^{4/3}}\left(
\frac ZA\right) ^{4/3}=\frac{492\rho _6^{1/3}}{g_{15}}\left( \frac ZA\right)
^{4/3}{\rm cm}.
\end{equation}
Fast magnetosonic waves propagate across the stellar crust without
attenuation if at any density $l>h$. The attenuation increases with
decreasing the matter density. Therefore, the maximum damping occurs in the
boundary region where magnetic field still does not suppress the electron
scattering. According to Eq. (\ref{d5}), in the region of maximum damping $%
\rho _6\approx 0.6H_{12}^{3/2}$ (the region is located several meters below
the surface). Then using Eqs. (\ref{d24}), (\ref{d241}) we obtain the
following condition at which waves pass the critical region and reach the
stellar surface without attenuation (for $_{26}^{56}$Fe lattice):
\begin{equation}
\label{d242}\left| \sin \theta \right| <\frac{0.004H_{12}^2g_{15}^{1/2}}{%
T_6^{5/2}}{\rm .}
\end{equation}
For $T=10^6$K, $H=4\times 10^{12}{\rm Gs}$ and $g=10^{15}{\rm cm/s}^2$ only
waves that in the critical region propagate along magnetic lines within the
angle $\left| \theta \right| <0.064{\rm rad}=3.7^{\circ }$ (that is only
about $0.2\%$ of the total vortex radiation) reach the stellar surface
without attenuation. However, with decrease of the surface temperature the
angle $\theta $ becomes larger, e.g., if $T=5\times 10^5$K, we obtain $%
\left| \theta \right| <21^{\circ }$. Due to anisotropic attenuation the
observed vortex radiation should be pulsed if magnetic poles are displaced
from the axis of star rotation (like in pulsars). The radiation is maximum
when the line of sight is close to the magnetic axis. According to Eq. (\ref
{d242}), for colder NSs the shape of vortex pulse becomes broader.

\section{Transformation of fast magnetosonic waves into electromagnetic
radiation at the stellar surface}

Let us consider the boundary between the stellar crust and atmosphere. One
can approximately treat the atmosphere as a vacuum because its density is
much less then density of the crust. When a magnetosonic wave incidents on
the boundary, it undergoes partial reflection and partially transforms into
electromagnetic wave which propagates into the vacuum. The relation between
the waves is determined by the boundary conditions at the surface of
separation, which require the tangential component of the electric field and
the normal component of the magnetic field to be equal. Conservation of the
energy flux imposes another boundary condition \cite{Land60}:
\begin{equation}
\label{bc}\left[ P+\frac{(H_t^2-H_n^2)}{8\pi }\right] =0,
\end{equation}
where $P=P_0+u_s^2\rho ^{\prime }$ is the pressure, $\rho ^{\prime }$ is the
density oscillation of electron quasiparticles, $H_t$ and $H_n$ are
tangential and normal components of the magnetic field. The square brackets
denote the difference between the values on the two sides of the surface.
From Eq. (\ref{bc}) we obtain the following condition for the oscillating
part of the magnetic field ${\bf h}$ and the density $\rho ^{\prime }$:
\begin{equation}
\label{bc1}\left[ u_s^2\rho ^{\prime }+\frac{H_th_t}{4\pi }\right] =0,\qquad
\left[ h_n\right] =0.
\end{equation}

The problem can be substantially simplified because near the stellar surface
the Alfven velocity is much larger then the speed of sound $u_A\gg u_s$. For
$H=10^{12}$Gs, $\rho _e=10^3(m_e/m_p){\rm g/cm}^3$ we obtain $u_A=4\times
10^{11}{\rm cm/s}\gg c$. In this case the second term in Eq. (\ref{bc1}) is
in the factor $u_A^2/u_s^2\gg 1$ greater than the pressure oscillation $%
u_s^2\rho ^{\prime }$ and the boundary condition reduces to $\left[
h_t\right] =0$. Also, under the condition $u_A\gg u_s$ the fast magnetosonic
wave propagates with the velocity $u=u_A/\sqrt{1+u_A^2/c^2}$ which is
independent of the wavevector ${\bf k}$ (see Eq. (\ref{m10})). The particle
velocity ${\bf V}$ in the wave lies in the $kH$ plane and perpendicular to $%
{\bf H}$, while oscillating magnetic ${\bf h}$ and electric fields ${\bf E}$
are perpendicular to each other and ${\bf k}$:
\begin{equation}
\label{eh}{\bf E}=\frac{u_A}{\sqrt{c^2+u_A^2}}({\bf h}\times \hat k),
\end{equation}
where $\hat k$ is a unit vector in the direction of ${\bf k}$ (see Fig. 3).
We obtain that the fast magnetosonic wave near the stellar surface is
similar to an electromagnetic wave in a medium with the dielectric constant $%
\varepsilon =1+c^2/u_A^2$ and the boundary conditions at the interface
coincide with those for electric and magnetic field at the boundary of the
dielectric medium. Therefore, transformation of the fast magnetosonic wave
into electromagnetic one is described by the same formulas as refraction of
electromagnetic waves at the interface between the dielectric and vacuum.

For example, let us consider the case when the magnetic field ${\bf H}$ lies
in the plane of incident wave. Then, the oscillating electric field ${\bf E}$
is perpendicular to the plane of incidence. The transmission coefficient is
given by
\begin{equation}
\label{tc}\tau =\frac{\sin (2\theta _1)\sin (2\theta _2)}{\sin ^2(\theta
_1+\theta _2)}{\bf ,}
\end{equation}
where $\theta _1$ is the angle of incidence and $\theta _2$ is the angle of
refraction ($\sin \theta _2=\sqrt{\varepsilon }\sin \theta _1$). For normal
incidence ($\theta _1=0$) the transmission coefficient is
\begin{equation}
\label{tc1}\tau =\frac{4\sqrt{\varepsilon }}{(\sqrt{\varepsilon }+1)^2}=%
\frac{4\sqrt{1+c^2/u_A^2}}{\left( \sqrt{1+c^2/u_A^2}+1\right) ^2}{\bf .}
\end{equation}
If $u_A\gg c$ (which is the case when $H>10^{11}{\rm Gs}$) the magnetosonic
wave is completely transformed into electromagnetic wave ($\tau \approx 1$).
In the opposite limit $u_A\ll c$ the transmission coefficient is small and $%
\tau $ $\approx 4u_A/c$. Fig. 4 shows the general scheme of the vortex
mechanism of NS radiation.

\section{Spectrum of neutron star radiation}

The spectral density of sound waves power radiated from the stellar core in
unit solid angle is given by Eq. (\ref{mss}). However, as we have shown,
only a fraction of magnetosonic waves that in the critical region propagate
in the direction close to the magnetic lines reaches the stellar surface and
only a fraction is transformed into electromagnetic waves which propagate
into surrounding space. The fraction depends, in particular, on magnetic
field strength and its distribution over the stellar surface.

Now let us discuss the region of applicability of Eq. (\ref{mss}). There is
a fundamental limitation that affects the spectrum of vortex radiation in
the infrared band. The limitation is that at any frequency the
electromagnetic radiation produced by vortices should not exceed black body
radiation produced by the surface $4\pi R^2$ with the same temperature $T$.
This statement is one of the formulations of Kirchhoff's law which is a
consequence of general thermodynamic principles. We assume that vortices are
excited thermally and if at some frequency the vortex radiation becomes
comparable with radiation of a black body this means that at such frequency
the rate of thermal excitation imposes the main restriction on the radiated
power: vortices radiate maximum power which can be pumped from the thermal
reservoir. Kirchhoff's law results in the following limitation on the
frequency at which Eq. (\ref{mss}) describes the spectrum of electromagnetic
radiation produced by vortices:
\begin{equation}
\label{m102}\frac \omega {2\pi }>\frac{0.24c^{6/5}m_n^{3/5}\Omega ^{2/5}}{%
\hbar ^{3/5}\ln ^{1/5}(\omega _c/\omega )},
\end{equation}
where $\omega _c=\hbar /4m_n\xi ^2\approx 10^{20}{\rm Hz}$. For $\Omega
/2\pi =4{\rm s}^{-1}$ we obtain $\omega /2\pi >1.5\times 10^{14}$Hz. At
lower frequencies the spectrum of vortex radiation (in unit solid angle)
follows Planck's formula
\begin{equation}
\label{m104}P_{{\rm v}}(\omega )=\frac{\hbar \omega ^3R^2}{4\pi ^2c^2\left[
\exp (\hbar \omega /k_BT)-1\right] }\approx \frac{\omega ^2R^2k_BT}{4\pi
^2c^2},
\end{equation}
where $T$ is the temperature of the NS core.

One should mention that Eq. (\ref{m102}) for the kink frequency contains no
free parameters. To take into account the gravitation redshift $z$ one
should divide Eq. (\ref{m102}) by $1+z$. Consequently, the observation of
the kink in the infrared band allows measurement of the gravitation redshift
at the core surface.

To compare our theory with available spectrum observations in different
frequency bands it is convenient to represent the spectral density of vortex
radiation $P_{{\rm v}}(\omega )$ as a power law $P_{{\rm v}}(\omega )\propto
\omega ^\alpha $, where $\alpha $ is the spectral index. The $\sqrt{\ln
(\omega _c/\omega )}$ function in the denominator of Eq. (\ref{mss}) shifts
the spectral index by a small value $1/2\ln (\omega _c/\omega )$. The shift
depends weakly on $\omega $ and changes the spectral index of vortex
radiation from $\alpha \approx -0.46$ in the optical band to $\alpha \approx
-0.43$ in the $X-$ray band.

Apart from vortex contribution there is thermal radiation from the NS
surface, which for middle-aged pulsars dominates in the ultraviolet and soft
$X-$ray bands. One should note that spectrum of thermal radiation may
substantially differ from a black body. E. g., this is the case for NS\
atmospheres composed from light elements (hydrogen, helium) \cite{Roma87}.
However, despite of the big difference, the observed thermal spectrum can be
well fitted with a hydrogen (helium) atmosphere model as well as with a
black body spectrum \cite{Yako99a}. The point is that spectrum of thermal
radiation is usually detected only in the soft $X-$ray band. No data are
available in the ultraviolet band due to strong interstellar absorption. In
the soft $X-$ray band one can successfully fit the data assuming two black
body components: hot polar caps and contribution from the rest (colder) NS
surface. In the case of hydrogen (helium) atmosphere the data are well
fitted with only one component contribution. In the optical-ultraviolet band
the two models predict substantially different spectra and the hydrogen
(helium) atmosphere curve lies above the black body giving the intensity
several times greater. However due to the lack of data in this spectral
region it is difficult to choose the adequate atmosphere model.

In this paper we estimate thermal radiation assuming helium atmosphere and
use the numerical results obtained by Romani \cite{Roma87}. The model fits
well the thermal spectrum with only one free parameter (the effective
surface temperature $T_{{\rm eff}}$). However, the main reason for choice of
helium atmosphere is that if there is vortex radiation from the stellar
surface then the atmosphere, in principle, can not be a black body because
black body surface layer would absorb all passing radiation. From this point
of view the detection of vortex radiation is a strong argument against black
body atmosphere models.

The total radiation from a NS is given by the sum of vortex (Eqs. (\ref{mss}%
), (\ref{m104})) and thermal components:
\begin{equation}
\label{m103}P(\omega )=aP_{{\rm v}}(\omega )+P_{{\rm th}}(\omega ),
\end{equation}
where $P_{{\rm th}}(\omega )$ is the spectrum of surface thermal radiation.
In Eq. (\ref{m103}) we introduced a dimensionless parameter $a$. The
parameter is useful when we compare theoretical predictions with radiation
observed from pulsars. The free parameter $a$ takes into account partial
absorption of the vortex radiation near the stellar surface, geometrical
effects related to unknown magnetic field distribution and position of the
line of site with respect to the cone swept by star's magnetic axis.
According to Eq. (\ref{d242}), for typical parameters of NSs and the surface
temperature $T_s\approx 10^6$K only about $0.2\%$ of vortex radiation
reaches the stellar surface and transforms into electromagnetic waves.
However, the outgoing radiation is concentrated in a small solid angle near
the magnetic axis which is determined by distribution of magnetic field. For
typical pulse shapes this angle is of the order of $0.1{\rm rad}$. That is
parameter $a$ should be of the order of $0.002\cdot 2\pi /0.1\approx 0.1$.
However, according to Eq. (\ref{d242}), with decreasing the surface
temperature the parameter $a$ increases as $a\propto 1/T_s^5$ and for colder
NSs becomes of the order of $1$.

In next section we compare the vortex radiation for different NSs with the
thermal surface contribution. In such relative comparison the neutron star
radius $R_s$ (which is approximately equal to the core radius $R$) cancels
in equations. As a result, in estimates$\,$there are only two free
parameters: the core temperature $T$ and the geometrical factor $a$. The
effective surface temperature $T_{{\rm eff}}$ is directly determined from
the thermal spectrum.

\section{Discussion}

Since the discovery of pulsars in 1967 numerous attempts were made to detect
radiation from a NS surface. Observations were mainly successful in soft $X-$%
ray band because of high thermal luminosity in this region. Recent
observations with ROSAT ($0.24-5.8\times 10^{17}$Hz), ASCA ($0.17-2.4\times
10^{18}$Hz) and Chandra obtained a detailed spectral information of a number
of pulsars at $X$-ray energies. Only a few pulsars were observed in optical
and ultraviolet bands and only some of them have detectable emission in
these bands. In Fig. 5 we compare the observed radiation spectrum of
middle-aged pulsars PSR B0656+14 ($\Omega /2\pi =2.6{\rm s}^{-1}$) and Vela (%
$\Omega /2\pi =11.2{\rm s}^{-1}$) with the radiation spectrum predicted by
our theory. Thermal radiation of the stellar surface dominates in the
ultraviolet and soft $X-$ray bands. We assume helium atmosphere with the
effective surface temperature $T_{{\rm eff}}=4.9\times 10^5$K for PSR
B0656+14 and $T_{{\rm eff}}=7.8\times 10^5$K for Vela. The NS radius $R_s$
is related to the pulsar distance $D$: $R_s/10{\rm km}=5.28\sqrt{1+z}(D/1%
{\rm kpc})$ for PSR B0656+14 and $R_s/10{\rm km}=4.1\sqrt{1+z}(D/1{\rm kpc})$
for Vela ($z$ is the NS redshift). For PSR B0656+14 and $D=0.25{\rm kpc}$ we
obtain $R_s\approx 13\sqrt{1+z}{\rm km}$, while for Vela and $D=0.35{\rm kpc}
$ the surface radius is $R_s\approx 14\sqrt{1+z}{\rm km}$. The vortex
contribution (dash line) dominates in the infrared, optical and hard $X-$ray
bands, where its spectrum has a slope $\alpha \approx -0.45$. In the far
infrared band vortices radiate the maximum power which can be pumped by
thermal excitation. In this band the radiation spectrum changes its behavior
and follows Planck's formula with $P(\omega )\propto \omega ^2$. The sum of
the vortex and thermal components is displayed with the solid line in Fig.
5. For PSR B0656+14 we take the core temperature $T=6.4\times 10^8$K and the
geometrical factor $a=0.18$, while for Vela $T=8\times 10^8$K and $a=0.33$.
The observed broad-band spectrum is consistent with our model for typical NS
parameters. This suggests that the vortex mechanism of radiation operates in
a broad frequency range from IR to hard $X-$rays. One should mention that
the measured IR-optical spectrum has nonmonotonic behavior, which probably
indicates the presence of unresolved spectral features or could be partially
due to contamination of the pulsar flux by the nearby extended objects \cite
{Kopt00}. The slope of the non-thermal component is less steep than it was
deduced by Pavlov et al. 1997 and Kurt et al. 1998 based on the optical-UV
data only. Also, in the optical band the radiation of middle-aged pulsars
was found to be highly pulsed with pulses similar to those in the hard $X-$%
ray band \cite{Shea98,Shea97}. This agrees with the vortex mechanism which
predicts pulsed radiation.

In Fig. 6 we plot the broadband spectrum of the Geminga pulsar ($\Omega
/2\pi =4.2{\rm s}^{-1}$) in optical$-X$-ray bands. Solid line is the fit by
the sum of the vortex and thermal contributions. In calculations we take $T_{%
{\rm eff}}=2\times 10^5$K, $T=3\times 10^8$K and $a=0.22$. The radius of the
Geminga pulsar is: $R_s/10{\rm km}=5.21\sqrt{1+z}(D/1{\rm kpc})$, which for $%
D=0.16{\rm kpc}$ gives $R_s\approx 8.3\sqrt{1+z}{\rm km}$. Our fit
reasonably agrees with the observed spectrum apart from the dips in the
IR-optical bands which probably appear due to ion cyclotron absorption \cite
{Mart98}. Also there is strong scattering of the data in the hard $X-$ray
band which makes difficult comparison with the theory and, hence, estimates
are less reliable. The data scattering might indicate on unresolved emission
and absorption lines in this spectral region.

It is worth to note that the formula for vortex radiation contains two free
parameters ($T$ and $a$). As a result we can not determine each of these
parameters separately from the observed fluxes, only their combination.
However, the spectral index of vortex radiation is a fixed parameter in our
theory. If a single mechanism of nonthermal radiation operates in the broad
range, then the continuation of the $X-$ray fit to the optical range
determines the spectral index of the nonthermal component with a big
accuracy. Good quantitative agreement of our theory with the observed
spectral index serves as a strong evidence that the vortex mechanism is
responsible for radiation of middle-aged NSs in the IR, optical and hard $X-$%
ray bands. Also, there is an evidence of vortex contribution in radiation of
young Crab pulsar. In optical band an emission during the ``bridge'' phase
interval has been detected from the Crab with the spectral index $\alpha
_{opt}=-0.44\pm 0.19$ \cite{Gold00}. The index differs from the flat
spectrum of the peaks and can indicate on the vortex component.

Moreover, according to Eq. (\ref{mss}), the vortex contribution
exponentially decreases at $\hbar \omega >k_BT$. Observation of such
spectrum behavior allows us to determine directly the temperature of the
stellar core $T$ and can be a possible test of our mechanism. For PSR
B0656+14 and the Vela pulsars the power law spectrum in the hard $X-$ray
band shows no changes up to the highest frequency $2\times 10^{18}$Hz at
which the data are available (see Fig. 5). According to our theory, this
indicates that temperature of the NSs core is larger than $2\times 10^8$K.
Measurements of spectra of middle-aged pulsars in the $10^{18}$-$10^{19}$Hz
range are needed to search for possible manifestations of the core
temperature.

One should note that recent observation of the Vela pulsar with the Rossi $%
X- $ray Timing Explorer has covered the energy band $2-30$keV ($%
0.49-7.3\times 10^{18}$Hz). These data in combination with OSSE observations
($70-570$keV) allows us to estimate the temperature of the Vela core. Light
curves of the Vela pulsar have several peaks. We associate the vortex
radiation with the second optical peak (see Fig. 1 in Harding {\rm et al.}
(1999) and Fig. 4 in Strickman {\rm et al.} (1999)). In the ROSAT band ($%
0.06-2.4$keV) the vortex peak is weakly distinguishable on the background of
strong thermal emission. However, the peak is clearly seen in the Rossi band
in which thermal radiation is negligible. The peak becomes very small in the
OSSE band and completely disappears at EGRET energies ($>100$MeV). The
vortex peak gives the main contribution to the average radiation intensity
in the optical and hard $X-$ray bands and its spectral index agrees with our
theory. One should note that in the Rossi band other peaks appear in the
Vela light curves \cite{Hard02}. Position of those peaks coincides with
pulses at EGRET energies and, therefore, they have a different origin. The
new peaks give the main contribution to the average radiation intensity at
energy greater then $10$keV. As a result, at such energies the average
intensity carries no information about vortex radiation and a light curves
study is necessary to separate the vortex contribution. To estimate the core
temperature one should trace the evolution of the vortex peak in the energy
interval $1$keV$-1$MeV. According to Eq. (\ref{mss}), the frequency
dependence of the peak amplitude depends only on the core temperature $T$.
To our knowledge such analysis of the vortex peak amplitude has not been
done yet. Nevertheless, it is possible to estimate the core temperature from
available data. If we assume that the amplitude of the vortex peak follows
Eq. (\ref{mss}), then we can find total (integrated over frequency)
radiation intensities of vortices in different energy bands. Then we compare
the detected integrated intensities of the vortex peak in the Rossi and OSSE
bands \cite{Stri99} with those predicted by Eq. (\ref{mss}). We found that
if there was no temperature decay of the vortex peak the total vortex
intensity in the OSSE band should be three times larger than those actually
observed. This indicates on the decay of the vortex spectrum and results in
the following estimate of the core temperature of the Vela pulsar: $T\approx
8\times 10^8$K.

The temperature estimate ($T>10^8$K) allows us to make a conclusion about
the interior constitution of the NSs. A hot NS cools mainly via neutrino
emission from its core. Neutrino emission rates, and hence the temperature
of the central part of a NS, depend on the properties of dense matter. If
the direct Urca process for nucleons ($n\rightarrow p+e+\bar \nu _e$, $%
p+e\rightarrow n+\nu _e$) is allowed, the characteristic cooling time is $%
\tau _{Urca}\sim 1\min /T_9^4$, where $T_9$ is the final temperature
measured in units of $10^9$K \cite{Peth92}. Such neutron star cools to $10^8$%
K in weeks. However, the characteristic age of PSR B0656+14 is about $10^5$%
yrs, while the Vela pulsar is about $10^4$ years old. So, the NS core cools
down to $10^8$K at least $10^6$ times slower then the rate predicted by the
direct Urca process. This estimate excludes equation of states incorporating
Bose condensations of pions or kaons and quark matter. The point is that the
neutrino emission processes for these states may be regarded as variants of
the direct Urca process for nucleons \cite{Peth92}. As a result, all these
states give rise to neutrino emission, though generally smaller, comparable
to that from the direct Urca process for nucleons which is inconsistent with
the discrepancy of the cooling rate in the factor $10^6$. The conclusion
about absence of exotic states of matter in the stellar core agrees with
recent measurement of gravitational redshift from a NS \cite{Cott02}.

It is worth to mention here that dissipation of rotation kinetic energy in
principle can produce heating of the stellar interior and supply energy for
neutrino emission \cite{Ripe95,Lars99}. Such possibility correlates with
recent observation of emission properties of 27 pulsars by Becker and
Tr\"umper (1997) which suggests that pulsars radiation is emitted at the
expense of their rotational energy (for $\Omega /2\pi =10{\rm s}^{-1}$ the
rotation energy $\sim M_{\odot }R_s^2\Omega ^2\approx 10^{50}{\rm erg}$).
The superfluid core possesses the main part of the rotation kinetic energy,
while radiation and dissipation of the angular momentum occurs from the
stellar surface. As a result, the outer crust slows down faster than the
superfluid core. The difference in the angular velocities between the
superfluid and normal components causes friction which heats the core and
transfers the angular momentum from the interior part to the outer crust.
Superfluid vortices can play a crucial role in such process \cite{Zhur01}.
One should note, however, that the neutrino emission rate comparable to that
from the direct Urca process also contradicts to the rotation energy loss of
NSs. Indeed, according to Lattimer {\rm et~al.} (1991), the neutrino
emissivity in the direct Urca process is $Q_{Urca}\sim 10^{-27}T^6{\rm %
erg/s\cdot cm}^3$ and the total power radiated by neutrino at $T=10^8$K is $%
P_{{\rm Urca}}=\frac 43\pi R^3Q_{{\rm Urca}}\sim 4\times 10^{39}{\rm erg/s}$%
. This value is much larger then the rate of the kinetic energy loss of the
middle-aged pulsars $10^{31}-10^{36}{\rm erg/s}$.

One should note that the characteristic time for cooling by the modified
Urca process for normal nucleons is $\tau _{{\rm modUrca}}\sim 1{\rm yr}%
/T_9^6$ \cite{Peth92}. That is if the direct Urca process cannot occur, the
core temperature will exceed $\sim 10^8$K for $\sim 10^6$ years. This result
does not contradict the estimate $T>10^8$K for middle-aged pulsars. However,
more accurate determination of the core temperature of the Vela pulsar ($%
T\approx 8\times 10^8$K) demonstrates that NSs cool down much more slowly
than could be explained by the modified Urca process for normal nucleons
(such process would cool the core down to $8\times 10^8$K during $4{\rm yrs}$%
, however the age of the Vela pulsar is about $10^4{\rm yrs}$). Possible
explanation of this observation is that superfluidities of neutrons and
protons in the stellar core are very strong so they substantially reduce all
neutrino processes involving nucleons. If $k_BT\ll \Delta $, where $\Delta $
is the gap in the excitation spectrum of a superfluid, the modified Urca
rates are reduced by a factor $\sim \exp (-2\Delta /k_BT)$. Using the
measured value of the core temperature one can estimate the critical
temperature of the superfluid transition. Theory predicts that in the
stellar core the critical temperature of the superconducting transition for
protons $T_{cp}$ is larger than the transition temperature of triplet
neutron superfluidity $T_{cn}$ \cite{Hoff70,Amun85a,Amun85b}. As a result,
the proton gap gives the main contribution to the suppression of neutrino
emission by the modified Urca process. The corresponding reduction factors
can be found in \cite{Yako01}. However, the appearance of neutron or proton
superfluid initiates an additional specific neutrino production mechanism
due to Cooper pair formation \cite{Yako99b}. Such mechanism of neutrino
emission dominates until it dies out at $T<0.1T_c$. The process involving
superfluid neutrons $n\rightarrow n+\nu +\bar \nu $ is more important in the
stellar core since their number greatly exceeds the number of protons. To
find $T_{cn}$ we assume $^3P_2$ triplet-state pairing of neutrons with the
gap $\Delta =\Delta _0\sqrt{1+3\cos ^2\theta }$. The neutrino emissivity due
to Cooper pairing of neutrons can be estimated using the result of Yakovlev
{\rm et al.} (1999b) $Q_{{\rm CP}}=2.2\times 10^{-42}T^7F(T/T_{cn}){\rm %
erg/s\cdot cm}^3$, where $F(x)\approx 1.27\exp (-2.376/x)/x^6$ for $x\ll 1$.
The stellar core cools down to the temperature $T$ during the time $\tau
=\int_T^\infty dTC(T)/Q_{{\rm CP}}(T)$, where $C(T)$ is the specific heat.
When $T\ll T_c$ the neutron (proton) contribution to the specific heat is
exponentially small. Therefore, $C(T)$ is determined by the electron
component $C(T)=(3\pi ^2)^{2/3}n^{2/3}k_B^2T/3c\hbar $, where $n\approx
10^{38}{\rm cm}^{-3}$ is the electron concentration in the stellar core.
Using parameters of the Vela pulsar $T\approx 8\times 10^8$K, $\tau \approx
10^4{\rm yrs,}$ we find the following critical temperature of triplet
neutron superfluidity in the center part of the Vela core $T_{cn}\approx
9\times 10^9$K. Again one should mention the possibility of heating the
stellar core by dissipation of the rotation energy. If such process is
responsible for the Vela temperature than in equilibrium the heating rate is
equal to the total power radiated by neutrino $(4/3)\pi R^3Q_{{\rm CP}}$.
The kinetic energy loss of the Vela pulsar is $6.9\times 10^{36}{\rm erg/s}$
\cite{Beck97}. If we assume that all kinetic energy loss goes in the
neutrino emission than we obtain $T_{cn}\approx 6\times 10^9$K ($R\approx 10%
{\rm km}$). The true value of $T_{cn}$ lies between this estimate and the
previous value $9\times 10^9$K found under the assumption of no extra
heating. More detailed analysis shows that the core temperature of
middle-aged pulsars is probably maintained by the heating due to the
dissipation of rotation energy. Our estimate of the critical temperature of
neutron superfluidity agrees well with realistic nucleon-nucleon interaction
potentials that fit recent world data for $pp$ and $np$ scattering up to $500%
{\rm Mev}$ \cite{Bald98}. Such potentials give $T_{cn}\approx 7.5\times 10^9$%
K at the density $2\psi _0$ that is expected at the center region of a
canonical neutron star without exotic inner core, and predict $T_{cn}\approx
2\times 10^9$K at the core-crust interface. One should mention that precise
knowledge about nucleon-nucleon interaction at high energies (up to $\sim 2.5%
{\rm Gev}$) is a result of progress in accelerator physics during the last
decade \cite{Arnd94,Arnd97}. Estimates of $T_{cn}$ made in 1970's and 1980's
usually substantially differ from the current value. E.g., Amundsen \&
Ostgaard (1985b) gave the maximum value of $T_{cn}$ in the stellar core $%
T_{cn}=9\times 10^8$K which is almost one order of magnitude less then the
current result. Recent experiments on nucleon-nucleon scattering favor in
strong superfluidity and superconductivity in the stellar core, such
observations are also supported by the theory \cite{Elga96,Bald01}.

It is worth to note that we may rule out the exotic states of matter only
for NSs which display vortex radiation in the hard $X-$ray band. However it
may not be the case for the whole NS population and a systematic study of
pulsars radiation in the $X-$ray band is necessary. If exotic states do
exist in some dense NSs the middle-aged pulsars with the core temperature $%
<10^7$K could be candidates for stars with Bose condensates or quark matter.

\section{Energy transport across the stellar crust}

The theory of NS cooling has been developing over more than 30 years. First
let us discuss the current understanding of temperature distribution in a
crust of middle-aged NSs. The question is important because NS cooling
theory is used as a method to determine the core temperature from the
surface temperature and, hence, to explore properties of superdense matter.
In current theories it is assumed that high density of the stellar interior
provides high thermal conductivity of internal layers. As a result, in about
$10^2-10^3$ years after the star formation, a wide, almost isothermal region
is formed within the entire core and the main fraction of the stellar crust
\cite{Yako99a}. The isothermal stellar interior is surrounded by a very thin
subphotospheric layer with thickness of the order of several meters. This
low dense surface layer produces thermal isolation of the NS interior. The
heat transport through the isolating layer determines the relation between
the surface temperature $T_s$ and the interior temperature $T$. In
particular, $T/T_s$ depends on the surface gravity and the chemical
composition of the outer NS envelope. For $T_s=5\times 10^5$K and a
nonmagnetized NS the ratio $T/T_s$ typically lies in the interval $20\div
100 $ \cite{Pote97}.

Possible presence of a strong magnetic field increases the uncertainty. The
point is that magnetic field suppresses the electron thermal conductivity
across the field lines and enhances along the lines. The effect of magnetic
field on NS cooling has been studied by several authors. E.g., a
relationship between the internal and surface temperatures in a NS with the
magnetic field normal to the surface was investigated by Van Riper (1988).
Page (1995) as well as Shibanov \& Yakovlev (1996) considered the NS cooling
with a dipole magnetic field.

As we have found based on spectra observations the effective surface
temperature of the Vela Pulsar is $7.8\times 10^5$K, while the core
temperature is about $8\times 10^8$K. This value of the core temperature is
much larger than those predicted in heat transport models discussed in
literature so far. The purpose of this section is to explain the big
difference. Here we show that if near the core surface the magnetic field is
parallel to the core it produces substantial heat isolation and can explain
the big difference between the core and surface temperatures. Such
assumption is simply a boundary condition for the magnetic field at the
superconductor boundary. In this model the main temperature drop occurs in a
thin layer near the crust bottom which differs from current cooling theories
that predict isothermal inner crust.

Let us consider heat transport across the stellar crust in the presence of a
strong magnetic field. We assume that free electrons give the main
contribution to thermal conductivity. In the presence of a strong magnetic
field ($E_F\gg \omega _{{\rm He}}\gg \nu _e$) the electron thermal
conductivity across the field can be estimated as
\begin{equation}
\label{d25}\kappa _{{\rm e}\perp }\approx \frac{\pi ^2n_ec^2k_B^2T\nu _e}{%
3E_F\omega _{{\rm He}}^2},
\end{equation}
where $\nu _e$ is the frequency of electron relaxation in momentum
directions which can be due to electron-phonon (Eq. (\ref{d231})) or
electron-electron collisions. At $T\ll T_F$, where $T_F$ is the Fermi
temperature, the frequency of electron-electron collisions for
ultrarelativistic electrons is
\begin{equation}
\label{d232}\nu _{{\rm e,e}}\sim \frac{e^4n_e^{1/3}}{\hbar ^2c}\left( \frac T%
{T_F}\right) ^2=3.2\times 10^{18}\frac{T_6^2}{n_e^{1/3}}{\rm s}^{-1}.
\end{equation}
For $n_e=10^{36}{\rm cm}^{-3}$, $T=10^8$K we obtain $\nu _{{\rm e,e}}\approx
3.2\times 10^{10}{\rm s}^{-1}$. Near the crust-core interface $\nu _{{\rm e,e%
}}>\nu _{2{\rm e,ph}}$ and, therefore, $\nu _{{\rm e,e}}$ should be
substituted into Eq. (\ref{d25}). As a result, the electron thermal
conductivity across the magnetic field is given by
\begin{equation}
\label{d26}\kappa _{{\rm e}\perp }\approx 8.5\times 10^{-29}\frac{n_eT_6^3}{%
H_{12}^2}\frac{{\rm erg}}{{\rm K\cdot cm\cdot s}}.
\end{equation}
For $n_e=10^{36}{\rm cm}^{-3}$, $T=10^8$K, $H=10^{12}{\rm Gs}$ we obtain $%
\kappa _{\perp }\approx 8.5\times 10^{13}{\rm erg/K\cdot cm\cdot s}$.
Magnetic field suppresses $\kappa _{{\rm e}\perp }$ in a factor $\nu _{{\rm %
e,e}}\nu _{1{\rm e,ph}}/\omega _{{\rm He}}^2\approx 10^{-11}$. One should
mention that Eq. (\ref{d26}) is valid when $\omega _{{\rm He}}^2\gg \nu _{%
{\rm e,e}}\nu _{1{\rm e,ph}}$. In the opposite limit the magnetic field does
not affect the thermal conduction.

Eq. (\ref{d26}) has to be compared with the phonon contribution to thermal
conductivity which for $T\ll T_D$ can be estimated as (for phonon relaxation
on electrons):
\begin{equation}
\label{d27}\kappa _{{\rm ph}}\approx \frac{4\pi ^3k_B^3T^2}{5\hbar ^2c}%
=1.97\times 10^9T_6^2\frac{{\rm erg}}{{\rm K\cdot cm\cdot s}}.
\end{equation}
For $T=10^8$K we obtain $\kappa _{{\rm ph}}\approx 2\times 10^{13}{\rm %
erg/K\cdot cm\cdot s}$. At such temperatures, the electron
thermoconductivity $\kappa _{{\rm e}\perp }$ is still larger than $\kappa _{%
{\rm ph}}$.

Now let us show that magnetic thermal isolation can substantially reduce the
heat transport from the stellar core. Under such condition the magnetosonic
waves produced by vortices can become the dominant mechanism of energy
transfer to the stellar surface. The superconducting core expels magnetic
field from its interior so that outside the core the magnetic lines are
parallel to the superconductor boundary. As a result, near the crust bottom
the magnetic field is tangential to the stellar core which produces
substantial thermal isolation. To estimate the heat transfer across the
magnetic layer we consider the following model. We assume that there is a
strong magnetic field parallel to the core surface in a thin layer of width $%
d$. The width $d$ is much less then the crust thickness ($d\ll 1{\rm km}$).
Outside the layer we assume the magnetic field to be weak or not tangential
to the surface so that outside the layer the field does not suppress thermal
conductivity across the crust.

One should also take into account that at high temperature the neutrino
emission processes result in substantial energy loss and can affect the
temperature profile in the inner crust. We assume that neutrino emission due
to Cooper pairing of neutrons is the main mechanism of neutrino generation
near the crust bottom. The process represents neutrino pair emission (of any
flavor) by a neutron whose dispersion relation contains an energy gap: $%
n\rightarrow n+\nu +\bar \nu $ \cite{Yako99b}. In this process the main
neutrino energy release takes place in the temperature range $0.2T_c\lesssim
T\lesssim 0.96T_c$, where $T_c$ is the critical temperature of neutron
superfluidity. At densities near the crust bottom $T_c\approx 10^9$K and,
therefore, the neutrino emission due to Cooper pairing of neutrons modifies
the temperature distribution at $2\times 10^8\lesssim T\lesssim 10^9$K. In
this temperature range we assume the following neutrino emissivity $Q\approx
10^{22}T_9^8{\rm erg/s\cdot cm}^3$ \cite{Yako99b}.

Fig. 7 shows the temperature distribution in the NS crust as a function of
the distance to the crust-core interface $z$. We take temperature at the
crust-core interface $T=8\times 10^8$K (the Vela pulsar), the width of
magnetic layer is $d=50{\rm m}$. Thermal conductivity across the layer is
given by Eq. (\ref{d26}) with $n_e=5\times 10^{36}{\rm cm}^{-3}$ and $%
H=4\times 10^{12}{\rm Gs}$. Neutrino emission substantially modifies the
temperature distribution and reduces the heat flux at $z<40{\rm m}$. At
larger distances from the core the temperature becomes low enough so that
neutrino emission is no longer important. In this region the heat flux is
conserved. At $z>d$ the thermal conductivity is not suppressed by magnetic
field and temperature remains approximately constant. Near the star surface
(at distance less then $1{\rm m}$ from the surface) there is another
temperature drop due to substantial decrease of thermal conductivity at low
densities. In estimates we take the crust width is equal to $1{\rm km}$.
Practically all heat flux from the stellar core is lost by neutrino emission
in a thin layer near the crust bottom. In our model the total energy losses
of the stellar core due to thermoconductivity is $7\times 10^{35}{\rm erg/s}$
(for core radius $R=10{\rm km}$), while only $1.4\times 10^{31}{\rm erg/s}$
passes the magnetic layer and contribute to thermal radiation from the
surface.

One should note that this value is about 20 times less then thermal energy
actually radiated from the Vela surface: for $R_s=10{\rm km}$, $T_{{\rm eff}%
}=7.8\times 10^5$K we obtain $4\pi R^2\sigma T_{{\rm eff}}^4=2.6\times
10^{32}{\rm erg/s}$. Such big discrepancy allows us to advance a hypothesis
that vortex radiation, rather then thermoconductivity, is the main mechanism
of heat transfer to the stellar surface. To check the hypothesis let us
estimate the temperature of a black body that radiates the same total power
as radiated by vortices from the superfluid core. Using Eq. (\ref{mss}) we
obtain
\begin{equation}
\label{d15}T_{{\rm eff}}=\frac{\hbar ^{1/4}c^{3/4}m_n^{3/8}\Omega
^{1/4}T^{3/8}}{k_B^{5/8}\ln ^{1/8}\left( \hbar \omega _c/k_BT\right) }.
\end{equation}
In our model the fast magnetosonic waves generated by vortices propagate
practically across the whole crust in a ballistic regime. For not too cold
NSs a thin layer near the stellar surface absorbs practically all
magnetosonic waves apart from a small fraction which reaches the surface. If
thermoconductivity across the crust is suppressed by magnetic field, then
vortex radiation can become the main mechanism of energy transfer from the
core to the surface. Under such condition the energy radiated by vortices
should be equal to the thermal energy radiated from the stellar surface and
Eq. (\ref{d15}) determines the effective surface temperature which is
directly measured from observations in the soft $X-$ray band. One should
note that Eq. (\ref{d15}) contains no free parameters because $T_{{\rm eff}}$
and $T$ can be measured independently and an uncertainty in $\omega _c=\hbar
/4m_n\xi ^2$ under the logarithm might change the answer only in a few
percents.

From observation of the Vela pulsar in the hard $X-$ray and soft $\gamma -$%
ray bands we estimated the Vela core temperature $T$ $\approx 8\times 10^8$%
K. For this value of the core temperature, $\Omega /2\pi =11.2{\rm s}^{-1}$
and $\omega _c/2\pi =10^{20}$Hz Eq. (\ref{d15}) gives $T_{{\rm eff}%
}=7.6\times 10^5$K, which approximately coincides with the effective surface
temperature of the Vela pulsar obtained from observations of thermal
spectrum in the soft $X-$ray band $T_{{\rm eff}}=7.8\times 10^5$K. Such
excellent agreement with observations is remarkable because the theory
contains no free parameters! This confirms our hypothesis that vortex
radiation, rather than thermoconductivity, is the main mechanism of heat
transfer to the stellar surface in middle-aged NSs (at least in some of
them). Also such agreement is a strong argument in favor of the theory of
vortex radiation.

As a result, one can use Eq. (\ref{d15}) to estimate the core temperature $T$
of middle-aged NSs from their surface temperature $T_{{\rm eff}}$ and obtain
the following practical relation:
\begin{equation}
\label{d16}T_8\ln ^{1/3}\left( 48/T_8\right) =\frac{0.22T_{{\rm eff}5}^{8/3}%
}{(\Omega /2\pi )^{2/3}}.
\end{equation}
For PSR B0656+14 $T_{{\rm eff}}=4.9\times 10^5$K, $\Omega /2\pi =2.6{\rm s}%
^{-1}$ and Eq. (\ref{d16}) gives the core temperature $T$ $\approx 6.4\times
10^8$K. For Geminga $T_{{\rm eff}}=2\times 10^5$K, $\Omega /2\pi =4.2{\rm s}%
^{-1}$and we obtain $T$ $\approx 0.3\times 10^8$K. One should mention that
for colder NSs (such as Geminga) absorption of vortex radiation near the
surface becomes small (see Eq. (\ref{d242})). In this case Eq. (\ref{d16})
is less reliable and can substantially underestimate the core temperature.
For colder NSs with still strong magnetic field the total power of vortex
radiation becomes larger than thermal surface emission.

\section{Conclusions and outlook}

In this paper we discuss a new possible mechanism of NS radiation produced
by vortices in the superfluid core. The key idea is that neutron star
interior is transparent for electron zero sound, the same way as it is
transparent for neutrinos. The waves of zero sound (in electron liquid) are
generated at the core-crust interface by superfluid vortices which undergo
thermal helical oscillations. Sound waves produced by vortices propagate
across the stellar crust towards the surface. In the presence of magnetic
field sound is coupled with electromagnetic waves. Virtually, they are
limiting cases of the same essence - the fast magnetosonic wave which is
known in magnetic hydrodynamics. Near the crust bottom, where the density is
large, the fast magnetosonic waves reduce to zero sound (the effect of
magnetic field is negligible). However, near the stellar surface the Alfven
velocity becomes larger than the speed of sound and the fast magnetosonic
waves behave as electromagnetic waves in a medium. As a result, at the
stellar surface the fast magnetosonic waves transform into electromagnetic
radiation in vacuum the same way as refraction of usual electromagnetic
waves at the dielectric-vacuum interface. For typical magnetic fields of the
middle-aged NSs the transformation coefficient is close to $1$.

Finite conductivity of the crust results in partial absorption of
magnetosonic waves in a thin layer near the stellar surface. As a result,
only a fraction of waves which in the absorbing layer propagate in the
direction close to magnetic lines reach the surface and transform into
electromagnetic radiation. The fraction depends on the magnetic field
strength and surface temperature. For typical parameters of middle aged NSs
the fraction is less than $1\%$, however for colder stars, such as Geminga,
it can be substantially larger (even of the order of $100\%$). As a
consequence of anisotropic attenuation and star rotation the observed vortex
radiation is pulsed, pulse shape is broader for colder stars. The part of
vortex radiation absorbed near the surface contributes to the heat transfer
from the core to the surface. Estimates show that at least for some NSs such
mechanism, rather than thermoconductivity, is the main source of heating the
stellar surface.

Stellar radiation produced by vortices has the spectral index $\alpha
\approx -0.45$ and for middle-aged pulsars dominates in the IR, optical and
hard $X-$ray bands. While the thermal surface radiation contributes mainly
in the ultraviolet and soft $X-$ray bands. The observed spectra of PSR
B0656+14, Vela and Geminga pulsars agree with our theory. In particular,
there is an excellent agreement with observations in two predictions where
the theory contains no free parameters: the spectral index of vortex
radiation $\alpha $ and the relation between the core and surface
temperatures (given by Eq. (\ref{d15})). There are also several predictions
which have to be confirmed by future observations. For example, the theory
predicts a kink of the vortex radiation spectrum in the infrared band (see
Eqs. (\ref{m102}), (\ref{m104}) and Figs. 5,6) and an exponential decay of
the spectrum in the hard $X-$ray band at $\hbar \omega >k_BT$ (see Eq. (\ref
{mss})) which was not measured in details yet.

The vortex radiation is produced in the stellar interior and contains
important information about properties of the NS core, in particular, it
allows direct determination of the core temperature. Comparing theory with
available spectra observations we found that the core temperature of the
Vela pulsar is $T\approx 8\times 10^8$K, which is the first measurement of
the temperature of a neutron star core. It is important to note that the
temperature we found is much larger than those predicted in heat transport
models discussed in literature so far. To understand the big difference we
consider heat transport in a model with magnetic field parallel to the core
surface (which is just a boundary condition at the superconductor boundary)
and show that the effect of magnetic thermal isolation can explain so high
value of the core temperature as compared with the stellar surface
temperature.

A direct measurement of the core temperature by means of detection the
vortex radiation opens a new perspective in study superdense matter. In
particular, our estimate of the interior temperatures of PSR B0656+14, Vela
and Geminga pulsars rules out possibility of exotic states, such as Bose
condensation of pions or kaons and quark matter, in these objects. Also it
allows us to extract the superfluid transition temperature for neutrons in
the center region of the stellar core $T_c=(7.5\pm 1.5)\times 10^9$K which
is the first determination of such parameter from observations.

Detection of vortex radiation opens a possibility to study composition of NS
crust (NS interior spectroscopy). Since magnetosonic waves generated by
vortices propagate through the stellar interior the spectrum of vortex
radiation should contain (redshifted) absorption lines which correspond to
low energy excitations of nuclei that form NS crust. E. g., the $_{26}^{57}$%
Fe nucleus has an excited state with the energy $14.4$kev$=3.5\times 10^{18}$%
Hz, which would produce an absorption line in the hard $X-$ray band of
vortex radiation spectrum if temperature of the stellar core is greater than
$1.7\times 10^8$K. The other possible absorption lines in the $X$ ray band
can be produced by nuclei $_{21}^{45}$Sc ($12.4$kev), $_{25}^{56}$Mn ($26$%
kev), $_{32}^{73}$Ge ($13.5$kev) or by heavy nuclei $_{92}^{235}$U ($13$%
kev), $_{97}^{249}$Bk ($8.8$kev). One should mention that bottom layers of
neutron star crust may contain exotic nuclei with mass number up to about $%
600$ and vortex radiation can be a possible tool to study their properties.
However, the absorption lines of nuclei are narrow and large spectral
resolution is necessary for their detection. If temperature of the stellar
crust $T_{crust}\sim 10^8$K the relative thermal line width is $\sqrt{%
2k_BT_{crust}/Mc^2}\approx 2\times 10^{-5}$. Under such condition the star
rotation gives the main contribution to the line broadening and the
corresponding Doppler line width for middle-aged pulsars can be estimated as
$\Omega R_s/c=0.001-0.01$.

Another challenging goal for future study is interaction of zero sound with
exotic states of matter which might exist in more dense NSs. The point is
that generation of zero sound by vortices is only one of the possible
mechanisms. If an exotic state of matter absorbs zero sound at some
frequency, then, according to Kirchhoff's law, it will radiate zero sound at
this frequency. As a result, spectrum of stellar radiation must contain
characteristic emission lines corresponding to such processes which allows a
direct spectroscopic study of superdense matter. In the limiting case when
the inner core matter absorbs all zero sound waves the stellar radiation
spectrum should be close to radiation of a black body with the temperature
of the core. One should look for such objects among bright galactic $X-$ray
sources. Also zero sound can be emitted in reactions between elementary
particles that constitute the stellar core. It is known that some pulsars
are powerful sources of $\gamma -$rays and it would be interesting to study
the connection between $\gamma $ radiation and zero sound. It is also
interesting to investigate a possible connection between electron sound and
nonthermal radiation of young (Crab like) pulsars which reveals flat
spectrum in the optical band.

\acknowledgements

I am very grateful to A. Fetter, G. Shlyapnikov, M. Binger, S. T. Chui, V.
Ryzhov for valuable discussions and the Aspen Center for Physics where part
of the results has been obtained. This work was supported by the National
Science Foundation, Grant No. DMR 99-71518 and by NASA, Grant No. NAG8-1427.

\appendix

\section{Zero sound in electron liquid}

In 1957 Landau predicted existence of collective sound excitations in Fermi
systems which exist in collisionless regime and called them ``zero sound''
\cite{Land57}. Zero sound waves propagate with the speed $u_s$ greater than
the Fermi velocity $v_F$ which excludes possibility of collisionless Landau
damping. Later, zero sound was observed in liquid $^3$He \cite{Keen65,Abel66}
and in electron liquid in metals \cite{Burm89,Bezu91,Bezu95}. The appearance
of zero-sound oscillations in charged Fermi liquids was studied by Silin
(1958) and Gor'kov \& Dzyaloshinskii (1963). In particular, Gor'kov \&
Dzyaloshinskii (1963) have shown that for weakly interacting electrons
ordinary zero sound can be propagated in the directions of the symmetry axes
in metals while spin zero sound apparently exists in any type of metal. As a
consequence of collective motion the attenuation of zero sound can be
atypically small compared to single particle excitations in Fermi liquids.
E.g., in normal $^3$He liquid under the pressure $27{\rm atm}$ the
attenuation length of longitudinal zero sound $2500$ times exceeds the free
path of single quasiparticles and the ratio substantially increases with
increasing the pressure \cite{Corr69}. For huge pressures, such as in a
neutron star crust, it is reasonably to expect very small attenuation of
collective modes.

Due to the presence of long-range Coulomb interactions, the Landau theory of
Fermi liquid is not directly applicable to electron gas in metals. However,
in electron gases the Coulomb interaction between quasiparticles is screened
by the gas polarizability and effectively becomes a short-range. It is
therefore conceivable that density waves of quasiparticles could arise
having a typical sound spectrum. If we take into account screening the
effective interaction potential between quasiparticles in momentum space is
given by
\begin{equation}
\label{at1}V_{{\rm eff}}(\omega ,{\bf k})=\frac{4\pi e^2}{k^2\varepsilon
_l(\omega ,{\bf k})},\quad
\end{equation}
where $\varepsilon _l(\omega ,{\bf k})$ is the longitudinal dielectric
function. In random phase approximation (RPA) \cite{Pine89}
\begin{equation}
\label{at2}\varepsilon _l(\omega ,{\bf k})=1-V(k)\chi ^0(\omega ,{\bf k}),
\end{equation}
where $V(k)=4\pi e^2/k^2$ is the interaction potential between bare
electrons and $\chi ^0(\omega ,{\bf k})$ is the density-density response
function for noninteracting fermions which is in the limit $k\ll k_F$, $%
\hbar \omega \ll E_F$ given by
\begin{equation}
\label{at3}\chi ^0(\omega ,{\bf k})=\frac{3n}{2E_F}\left( \frac \omega {2v_Fk%
}\ln \left( \frac{\omega +v_Fk}{\omega -v_Fk}\right) -1\right) ,
\end{equation}
$n$ is the electron concentration, $v_F$ is the Fermi velocity. In this
section we assume absence of external magnetic field. Hence, in RPA
\begin{equation}
\label{at4}V_{{\rm eff}}(\omega ,{\bf k})=\frac{4\pi e^2}{\left[
k^2+q_{TF}^2\left( 1-\frac s2\ln \left( \frac{s+1}{s-1}\right) \right)
\right] },\quad
\end{equation}
where $s=\omega /v_Fk$, $q_{TF}^2=6\pi ne^2/E_F$ is the Thomas-Fermi
screening wavevector. Such form of the interaction potential suggests
screening of the electric charge. E.g., if $V_{{\rm eff}}(k)=4\pi
e^2/(k^2+q_c^2)$ the interaction potential in coordinate space has the form $%
V_{{\rm eff}}(r)=(e^2/r)\exp (-q_cr)$, that is the effective quasiparticle
interaction is short-range.

The RPA takes into account electron interaction only to the extent required
to produce the screening field. The effects arising from interaction between
screened charges (such as zero sound), are, therefore, beyond the RPA. One
should note, however, that screening depends substantially on $\omega $ and $%
k$. For $\omega \gg v_Fk$ ($s\gg 1$) the screening is small $V_{{\rm eff}%
}(\omega ,{\bf k})\approx 4\pi e^2/(k^2-q_{TF}^2/3s^2)\approx V(k)$ and,
hence, RPA is a valid theory. This is the reason why RPA is successful in
describing plasmons in metals for which $\omega \gg v_Fk$. Indeed, the
dispersion equation $\varepsilon _l(\omega ,{\bf k})=0$ with $\varepsilon
_l(\omega ,{\bf k})$ given by (\ref{at2}) has a solution $\omega ^2=4\pi
e^2n/m_e$ corresponding to oscillations with plasma frequency.

However, in the region where zero sound branch is expected (for weakly
interacting fermions this region is $s-1\ll 1$) Eq. (\ref{at4}) predicts
large screening and RPA can not capture collective modes in this region.
Instead, one can apply a phenomenological Landau theory based on the
effective screened interactions between quasiparticles. The Landau theory
describes interaction by means of dimensionless phenomenological parameters $%
F_l$, $G_l$ which measure the interaction strength as compared to the
kinetic energy. The parameters $F_l$, $G_l$ describe respectively the spin
symmetric and spin antisymmetric parts of the quasiparticle interaction. For
weakly interacting Fermi gas only the first two parameters $F_0$ and $G_0$
are relevant, they are related to each other by $G_0=-F_0$ which is the
property of exchange interaction between electrons. Parameters $F_0$ and $%
G_0 $ describe propagation of axisymmetric (longitudinal) zero sound.
Propagation of zero sound is possible when $F_0>0$ or $G_0>0$. The former
case corresponds to an ordinary zero sound, while the latter one is known as
spin zero sound in which oscillation of the quasiparticle distribution
function depends on the spin direction. Loosely speaking $F_0$ describes
propagation through the medium of a particle-hole pair with the total spin $%
0 $, while in the spin zero sound the total spin of the propagating
particle-hole pair is equal to $1$ \cite{Gott60}. The speed of ordinary zero
sound $u_s=sv_F$ is determined from the equation \cite{Land57}
\begin{equation}
\label{at5}\frac s2\ln \left( \frac{s+1}{s-1}\right) -1=\frac 1{F_0},\quad
\end{equation}
which for weakly interacting fermions $F_0\ll 1$ reduces to
\begin{equation}
\label{at6}s-1\approx 2\exp \left( -2-\frac 2{F_0}\right) .
\end{equation}
That is the speed of zero sound is exponentially close to $v_F$. The speed
of spin zero sound is described by the same Eqs. (\ref{at5}), (\ref{at6})
with $G_0$ instead of $F_0$. The property $G_0=-F_0$ suggests that one of
these parameters is positive. This is a manifestation of a general statement
that ordinary or spin longitudinal zero sound exist in any stable Fermi
liquid \cite{Merm67}.

For a short-range weak interaction between fermions the Landau parameters
can be expressed in terms of the interaction potential as $F_0=3nV_{{\rm eff}%
}(k=0)/2E_F$. Formal application of Eq. (\ref{at4}) at $k\rightarrow 0$, $%
s\rightarrow 1$ yields $F_0=2/\ln [(s-1)/2]$ $<0$, $|F_0|\ll 1$. This gives
us a hint that $F_0$ has a small negative value. The magnitude of $F_0$ can
be approximately estimated as the ratio of the Coulomb energy to the kinetic
(Fermi) energy. For relativistic electron gas in the stellar crust we
obtain: $F_0\approx -e^2k_F/E_F=-1/\alpha =-0.007$, where $\alpha =c\hbar
/e^2$ is the fine structure constant. Negative value of $F_0$ suggests that $%
G_0$ is positive ($G_0=-F_0\approx 0.007$) and, hence, the spin zero sound
can propagate through the stellar crust.

Such conclusion can be justified by consideration of the density-density
response function, for weakly interacting Fermi gas it is given by \cite
{Gott60}
\begin{equation}
\label{drf}\chi _\lambda (\omega ,{\bf k})=\frac{\chi ^0(\omega ,{\bf k})}{%
1-(-1)^\lambda V_{{\rm eff}}(\omega ,{\bf k})\chi ^0(\omega ,{\bf k})},\quad
\end{equation}
where $\lambda =0,1$ it the total spin of the particle-hole pair which can
be excited in a singlet ($\lambda =0$) or a triplet state ($\lambda =1$).
Collective modes of the system are determined by the poles of $\chi _\lambda
(\omega ,{\bf k})$. If we take $V_{{\rm eff}}(\omega ,{\bf k})$ from Eq. (%
\ref{at4}) then $\chi _\lambda (\omega ,{\bf k})$ has a pole at $\omega
=v_Fk $ for $\lambda =1$. This pole describes spin zero sound, while the
plasmon branch corresponds to a pole for $\lambda =0$. In such approach we
find that speed of zero sound is exactly equal to $v_F$ which corresponds to
the limiting case of noninteracting quasiparticles. The result formally
appears because $V_{{\rm eff}}(\omega ,{\bf k})$ vanishes at $s=1$ which
means infinite screening of Coulomb interaction. However, in real systems
there is no screening at distances less than the interparticle spacing $%
\approx 1/k_F$, that is instead of zero one should take $|V_{{\rm eff}%
}|\approx e^2k_F/n$. As a result, the position of the pole is shifted from $%
s=1$ and determined by the same Landau expression (\ref{at5}) with $%
G_0\approx e^2k_F/E_F$. In Appendix B we discuss zero sound from the
viewpoint of two component (particle-hole) picture of excitations in
electron liquid.

A remarkable feature of zero sound in a weakly interacting Fermi gas is that
its attenuation is exponentially small. Attenuation of zero sound occurs by
means of excitation of two electron-hole pairs, since single particle
processes (Landau damping) are not allowed ($u_s>v_F$). Gottfried and Pi\v c%
man (1960) have solved the problem about zero sound attenuation in a weakly
interacting Fermi gas for $T=0$ and obtained the following result for the
imaginary part of the sound frequency $^{}$%
\begin{equation}
\label{at7}{\rm Im}\omega =(s-1)\frac{\pi ^4\hbar \omega ^2}{2E_F}.\quad
\end{equation}
For weak interactions the multiple $s-1$ is exponentially small which
provides exponentially small sound attenuation. Formal reason for this is
the logarithmic singularity in the density-density response function for
noninteracting fermions $\chi ^0(\omega ,{\bf k})$ at $\omega =v_Fk$ (see
Eq. (\ref{at3})). Finite temperature can be immediately included using the
Landau formula \cite{Land57}, finally we obtain the free path length for
zero sound
\begin{equation}
\label{at8}l=\frac{u_s}{{\rm Im}\omega }=\frac{\hbar v_FE_F\exp \left(
2+2/G_0\right) }{\pi ^4\left[ (\hbar \omega )^2+(2\pi k_BT)^2\right] }.\quad
\end{equation}
For $G_0=0.007\,$ the exponential factor gives a multiple $10^{125}$, hence
zero sound propagates without attenuation across the dense stellar crust for
any imaginable values of frequency and temperature at which the Fermi liquid
theory is still valid ($\hbar \omega ,k_BT\ll E_F$). However, close to the
stellar surface the electron gas becomes non relativistic. In this region
the parameter $G_0$ depends on the density $G_0\approx
e^2k_F/E_F=0.65/a_Bn^{1/3}$, where $a_B=\hbar ^2/m_ee^2=0.53\AA $ is the
Bohr radius, and could be of the order of $1$ near the surface. However, in
the surface region the effect of magnetic field becomes substantial and
prevents the wave damping. We discuss this in another section.

\section{Equations of magnetic hydrodynamics for electron motion}

In this section we consider equations which describe wave propagation across
the stellar crust. Typical temperature in the crust of middle-aged NSs is
less then the melting temperature \cite{Pote97}. As a result the matter
forms a Coulomb crystal. One should note that zero-point ion vibrations do
not destroy the lattice made of heavy nuclei. The amplitude of zero-point
vibrations is commonly much smaller than typical inter-ion spacing. With
increasing density the relative amplitude becomes larger, so that the
vibrations can prevent crystallization at high $\rho $. The crystallization
is completely suppressed in the so called quantum liquids, which exist at $%
\rho _6>0.006A^4Z^6$ \cite{Jone96}. The suppression is pronounced for light
nuclei, however it does not occur in the lattice composed from heavy
elements which is likely the case in the stellar crust. E.g., for iron
lattice ($A=56$, $Z=26$) the density at which zero-point vibrations melt the
lattice is $\rho >2\times 10^{19}{\rm g/cm}^3$. This value is much larger
than the density at the crust bottom ($10^{14}{\rm g/cm}^3$) and, therefore,
vibrations do not destroy the lattice in the crust.

In our model we assume that crust consists of nuclei which form a lattice
and a gas of free electrons. Free neutrons appear in the inner crust.
Frequencies of radiation we are interested in are much larger than frequency
of electron-electron and electron-phonon collisions. Hence, the electron gas
can be treated as collisionless. As we have seen in the Appendix A, in
collisionless regime zero sound is a possible collective excitation in
electron liquid without magnetic field. It is known that in a strong
magnetic field, such that $u_A\gg v_F$ (here $u_A=H/\sqrt{4\pi \rho _e}$ is
the Alfven velocity for electrons, $\rho _e$ is the electron density), the
collective excitations of electrons in compensated metals reduce to
magnetoplasma waves \cite{Buch61,Kane63}. Their nature is similar to
magnetosonic and Alfven waves in magnetic hydrodynamics. The dispersion
relations of magnetoplasma waves can be obtained from the dielectric
permittivity of collisionless plasma in a magnetic field. Indeed, if only
electrons contribute to the permittivity then \cite{Lifs79}
\begin{equation}
\label{t0}\varepsilon _{\perp }=1-\frac{\omega _{pe}^2}{\omega ^2-\omega
_{He}^2},
\end{equation}
where $\omega _{pe}=\sqrt{4\pi e^2n_e/m_e}$ is the electron plasma frequency
and $\omega _{He}=eH/m_ec$ is the electron cyclotron frequency. For this $%
\varepsilon _{\perp }$ the dispersion equation for electromagnetic waves at $%
\omega \ll \omega _{He}$ has two solutions $\omega =ku_A/\sqrt{1+u_A^2/c^2}$
and $\omega =ku_A\cos \vartheta /\sqrt{1+u_A^2/c^2}$ describing the fast
magnetosonic and Alfven waves respectively (here $\vartheta $ is the angle
between the wavevector ${\bf k}$ and ${\bf H}$).

In magnetic hydrodynamics the fast magnetosonic wave reduces to usual sound
in the limit of small magnetic field (when $u_A\ll u_s$). Here we
demonstrate that such situation remains also valid in the collisionless
regime with zero sound playing the role of usual sound. For this purpose it
is convenient to use a definition of elementary excitations in electron
liquid in terms of electron-hole pairs. In such picture the excitation with $%
p>p_F$ (particle) has the energy $\varepsilon =v_F(p-p_F)$, while at $p<p_F$
(hole) the energy is $\varepsilon =v_F(p_F-p)$, where $p_F$ is the Fermi
momentum. In such definition the excitation energy is positive and gives the
energy excess above the ground state. Elementary excitations appear in pairs
so that the total number of particles is always equal to the total number of
holes. The dynamics of excitations is described by the collisionless kinetic
equation for the quasiparticle distribution function $n(t,{\bf r},{\bf p})$:

\begin{equation}
\label{t1}\frac{\partial n}{\partial t}+{\bf v}\frac{\partial n}{\partial
{\bf r}}-\frac{\partial n}{\partial {\bf p}}\int f({\bf p},{\bf p}^{\prime })%
\frac{\partial n({\bf p}^{\prime })}{\partial {\bf r}}\frac{d^3p^{\prime }}{%
(2\pi \hbar )^3}+e\left[ {\bf E+}\frac 1c\left( {\bf v\times H}\right)
\right] \frac{\partial n}{\partial {\bf p}}=0,
\end{equation}
where ${\bf v=}$ $\partial \varepsilon /\partial {\bf p}$ and ${\bf p}$ are
the quasiparticle velocity and momentum. The quasiparticle charge and the
mass are given by $e=|e|{\rm sign}(p_F-p)$ and $m=m_e{\rm sign}(p-p_F)$
respectively. The function $f({\bf p},{\bf p}^{\prime })$ describes
interactions between quasiparticles and determines the velocity of zero
sound. For simplicity we assume that $f({\bf p},{\bf p}^{\prime })$ does not
contain a dependence on the quasiparticle spins. The equilibrium
distribution function is $n_0=1/[\exp (\varepsilon /k_BT)+1]$, so that the
equilibrium quasiparticle number depends on temperature.

Let us first consider the case of zero external magnetic field. In general,
the particle-particle interaction differs from the interaction between a
particle and a hole. As a result, it is convenient to treat our system as
composed of two components. Then dynamics of the particle ($n_1$) and the
hole ($n_2$) distribution functions is described by the following equations

\begin{equation}
\label{z1}\frac{\partial n_1}{\partial t}+{\bf v}\frac{\partial n_1}{%
\partial {\bf r}}-\frac{\partial n_1}{\partial {\bf p}}\int \left[ f_{11}(%
{\bf p},{\bf p}^{\prime })\frac{\partial n_1({\bf p}^{\prime })}{\partial
{\bf r}}+f_{12}({\bf p},{\bf p}^{\prime })\frac{\partial n_2({\bf p}^{\prime
})}{\partial {\bf r}}\right] \frac{d^3p^{\prime }}{(2\pi \hbar )^3}-|e|{\bf E%
}\frac{\partial n_1}{\partial {\bf p}}=0,
\end{equation}
\begin{equation}
\label{z2}\frac{\partial n_2}{\partial t}+{\bf v}\frac{\partial n_2}{%
\partial {\bf r}}-\frac{\partial n_2}{\partial {\bf p}}\int \left[ f_{22}(%
{\bf p},{\bf p}^{\prime })\frac{\partial n_2({\bf p}^{\prime })}{\partial
{\bf r}}+f_{21}({\bf p},{\bf p}^{\prime })\frac{\partial n_1({\bf p}^{\prime
})}{\partial {\bf r}}\right] \frac{d^3p^{\prime }}{(2\pi \hbar )^3}+|e|{\bf E%
}\frac{\partial n_2}{\partial {\bf p}}=0.
\end{equation}
One can seek a solution of these equations in the form $n_\alpha =n_{0\alpha
}+\delta n_\alpha $, where $\delta n_\alpha =(\partial n_{0\alpha }/\partial
\varepsilon )\nu _\alpha ({\bf p})\exp [i({\bf kr}-\omega t)]$ is a small
correction, $\alpha =1,2$. Here the function $\nu _\alpha $ depends only on
the direction of ${\bf p}$ since $\partial n_0/\partial \varepsilon $ is
nonnegligible only when $p\approx p_F$. For simplicity we assume $f_{\alpha
\beta }({\bf p},{\bf p}^{\prime })$ to be constant. Then using $\partial
n_0/\partial {\bf p=v}\partial n_0/\partial \varepsilon $ and $%
d^3p=p_F^2d\Omega dp$ we obtain equations for $\nu _\alpha $
\begin{equation}
\label{z3}(\omega -{\bf kv)}\nu _1({\bf p})-{\bf kv}\int \left[ F_{11}\nu _1(%
{\bf p}^{\prime })+F_{12}\nu _2({\bf p}^{\prime })\right] \frac{d\Omega
^{\prime }}{4\pi }{\bf -}i|e|{\bf Ev}=0,
\end{equation}
\begin{equation}
\label{z4}(\omega -{\bf kv)}\nu _2({\bf p})-{\bf kv}\int \left[ F_{22}\nu _2(%
{\bf p}^{\prime })+F_{21}\nu _1({\bf p}^{\prime })\right] \frac{d\Omega
^{\prime }}{4\pi }{\bf +}i|e|{\bf Ev}=0,
\end{equation}
where $F_{\alpha \beta }=-4\pi p_F^2\int f_{\alpha \beta }(\partial
n_{0\beta }/\partial \varepsilon )dp/(2\pi \hbar )^3$ are dimensionless
Landau parameters and $d\Omega ^{\prime }$ means integration over directions
of ${\bf p}^{\prime }$. Eqs. (\ref{z3}), (\ref{z4}) have to be solved
simultaneously with the Maxwell equations for the electric field ${\bf E}$.
However, as we have mentioned in Appendix A, in the zero sound regime the
electric field of quasiparticles is screened by the rest part of electrons
which fill the Fermi sphere and, hence, in Eqs. (\ref{z3}), (\ref{z4}) one
can put ${\bf E=0}$. The screening effect is included into Landau parameters
$F_{\alpha \beta }$. One can consider the quasiparticles as moving in a
medium with a very large dielectric constant $\varepsilon (\omega ,{\bf k})$%
, then the Maxwell equation ${\rm div}(\varepsilon {\bf E})=4\pi \rho $
results in ${\bf E\cdot k}=0$ which justifies the omitting ${\bf E}$. In our
case of the two component particle-hole system $F_{11}=F_{22}$, $%
F_{12}=F_{21}$. Under such conditions Eqs. (\ref{z3}), (\ref{z4}) have the
following solutions describing longitudinal zero sound
\begin{equation}
\label{z5}\nu _1=\frac{C\cos \theta }{s-\cos \theta },\quad \nu _2=\pm \frac{%
C\cos \theta }{s+\cos \theta },
\end{equation}
where $s=\omega /kv_F=u_s/v_F$ and $\theta $ is the angle between the
wavevector ${\bf k}$ and ${\bf p}$. The equation for $s$ formally coincides
with the expression obtained by Landau for a neutral Fermi liquid
\begin{equation}
\label{z6}\frac s2\ln \left( \frac{s+1}{s-1}\right) -1=\frac 1{F_{11}\mp
F_{12}}
\end{equation}
in which the Fermi liquid interaction parameter $F=F_{11}\mp F_{12}$ \cite
{Duni72}. For the solution with the upper sign the particles and holes move
with the same average velocity and the total current is equal to zero. Such
solution also exists if there is no screening since the particle-hole motion
does not produce electric field. Unscreened Coulomb interaction between
particles is repulsive, that is $F_{11}>0$, while particle-hole interaction
is attractive $F_{12}<0$. This suggests that $F=F_{11}-F_{12}>0$ and, hence,
the zero sound branch always exists in the system (not subject to Landau
damping). For the solution with the lower sign the particles and holes move
in opposite direction producing quasiparticle current and density
oscillation. Such solution is allowed only when there is screening of the
electric field.

Now let us discuss the effect of magnetic field. Here it is convenient to
consider each of the components separately. For definiteness we derive
equations of magnetic hydrodynamics in terms of (quasi)particle motion and
use Eq. (\ref{z1}) with the additional magnetic term. The average
quasiparticle concentration $n_q$, the quasiparticle current density ${\bf j}%
_q$ and the average velocity ${\bf V}_q$ are given by
\begin{equation}
\label{t2}n_q=\int n_1d^3p,\quad {\bf j}_q=\int e{\bf v}n_1d^3p,\quad {\bf V}%
_q{\bf =}\frac{\int {\bf v}n_1d^3p}{\int n_1d^3p}.
\end{equation}
Integrating Eq. (\ref{z1}) over $d^3p$ we obtain the continuity equation
\begin{equation}
\label{t3}\frac{\partial n_q}{\partial t}+\nabla (n_q{\bf V}_q{\bf )}=0.
\end{equation}
To derive an equation of momentum conservation in collisionless regime one
should make an assumption about deformation in the distribution function.
The assumption is necessary only when we integrate the term ${\bf v}\partial
n_1/\partial {\bf r}$ in the kinetic equation. Such a term is nonnegligible
as compared to the magnetic field contribution only in the zero sound
regime. Hence, without loss of generality one can assume that perturbation
in the distribution function is similar to those for the longitudinal zero
sound. Deviation of the sound speed $u_s$ from $v_F$ is determined by
interactions between electrons $f$. However, at large densities the
interaction energy is small compared with the electron kinetic energy and,
therefore, $u_s\approx v_F$, that is $s-1\ll 1$. Under such condition,
according to Eq. (\ref{z5}), the deviation of the distribution function $%
\delta n_1$ from its equilibrium shape is extremely anisotropic. As a
result, one may assume that $\delta n_1$ is nonzero only in a small vicinity
of the direction given by the vector $\partial n_1/\partial {\bf r}$. Also
one can omit the small terms with $f$ describing interactions and take $n_0$
instead of $n_1$ in the term with the electric field ${\bf E}$. Further, we
assume that $\omega \ll \omega _{He}$. Such assumption makes possible to
omit the term with electric field in the kinetic equation. The point is that
the magnetic field ${\bf H}$ does not allow the electron to move in a
direction perpendicular to ${\bf H}$ and change its average momentum in this
direction under the influence of ${\bf E}$. So, if we are interested in the
electron motion on a time scale larger than the Larmor period (that is $%
\omega \ll \omega _{He}$) only component of ${\bf E}$ along ${\bf H}$ can
contribute to the change in electron momentum. Formally this follows from an
equality ${\bf E(}\partial n_0/\partial {\bf p)=Ev(}\partial n_0/\partial
\varepsilon {\bf )}$ in which the electron velocity ${\bf v=}\partial
\varepsilon /\partial {\bf p}$ after averaging over fast Larmor precession
has nonzero component only along ${\bf H}$. Therefore, in the kinetic
equation one can substitute ${\bf E}$ by its component along the magnetic
field. For simplicity of derivations we omit completely the term with the
electric field bearing in mind that solutions of final equations are valid
when they posses a property ${\bf E\cdot H=0}$, which is the case for
magnetosonic waves we are focusing on. Then multiplying Eq. (\ref{z1}) on $%
{\bf v}$ and integrating over $d^3p$ we obtain the following analog of
Euler's equation in collisionless regime
\begin{equation}
\label{t4}\frac{\partial (n_q{\bf V}_q{\bf )}}{\partial t}+v_F^2\nabla n_q-%
\frac 1{cm_e}({\bf j}_q{\bf \times H)}=0.
\end{equation}
Further we note that if quasiparticles move with an average velocity ${\bf V}%
_q$ then such motion induces an electric field ${\bf E}=-(1/c)({\bf V}_q{\bf %
\times H)}$. The reason is the large electric conductivity $\sigma $ of the
system which requires the total force exerting on quasiparticles to be equal
to zero. The induced electric field also acts on the rest part of electrons
filling the Fermi sphere. As a result, these electrons must move with some
average velocity ${\bf V}$ to insure the equality ${\bf E}+(1/c)({\bf %
V\times H)\approx 0}$. We obtain that magnetic field couples the motion of
quasiparticles and electrons inside the Fermi surface. Equation $({\bf V}_q%
{\bf \times H})\approx ({\bf V\times H})$ suggests that the quasiparticle
current ${\bf j}_q$ is related to the entire current density ${\bf j}$ by $(%
{\bf j}_q{\bf \times H)}=(n_q/n_e)({\bf j\times H)}$, where $n_e$ is the
total concentration of electrons in the system.

Further, using Maxwell's equations ${\rm curl}{\bf H=}4\pi {\bf j}%
/c+(1/c)\partial {\bf E}/\partial t$, ${\rm curl}{\bf E=}-(1/c)\partial {\bf %
H}/\partial t$ and ${\bf j=}\sigma [{\bf E}+(1/c)({\bf V}_q{\bf \times H)}]$
we obtain equations of magnetic hydrodynamics for quasiparticle motion (in
linearized form):
\begin{equation}
\label{t5}\frac{\partial {\bf H}}{\partial t}={\rm curl}({\bf V}_q\times
{\bf H})+\frac{c^2}{4\pi \sigma }\left( \Delta {\bf H-}\frac 1{c^2}\frac{%
\partial ^2{\bf H}}{\partial t^2}\right) ,
\end{equation}
\begin{equation}
\label{t6}\frac{\partial {\bf V}_q}{\partial t}=-\frac{v_F^2}{n_q}\nabla n_q-%
\frac 1{4\pi \rho _e}({\bf H}\times {\rm curl}{\bf H})-\frac 1{4\pi c^2\rho
_e}\left( {\bf H}\times \left( \frac{\partial {\bf V}_q}{\partial t}\times
{\bf H}\right) \right) +\frac{|e|}{4\pi cm_e\sigma }\left( \frac{\partial
{\bf V}_q}{\partial t}\times {\bf H}\right) ,
\end{equation}
where $\rho _e=m_en_e$ is the total density of electrons. The last term in
Eq. (\ref{t5}) describes attenuation of magnetosonic waves due to finite
conductivity $\sigma $. It is interesting to note, that for waves
propagating with velocity close to the speed of light (which is the case for
ultrarelativistic electrons in the stellar crust) the dissipative term goes
to zero and attenuation of magnetosonic waves is substantially suppressed.

In the limit of large conductivity ($\sigma \rightarrow \infty $) Eqs. (\ref
{t3}), (\ref{t5}), (\ref{t6}) reduce to usual equations of nondissipative
magnetic hydrodynamics in which speed of sound is equal to the Fermi
velocity and quasiparticle density undergoes oscillations instead of the
total density of electrons:
\begin{equation}
\label{m1}\frac{\partial {\bf h}}{\partial t}={\rm curl}({\bf V}_q\times
{\bf H}),
\end{equation}
\begin{equation}
\label{m2}\frac{\partial n_q^{\prime }}{\partial t}+n_q{\rm div}{\bf V}_q=0,
\end{equation}
\begin{equation}
\label{m3}\frac{\partial {\bf V}_q}{\partial t}=-\frac{v_F^2}{n_q}\nabla
n_q^{\prime }-\frac 1{4\pi \rho _e}({\bf H}\times {\rm curl}{\bf h})-\frac 1{%
4\pi c^2\rho _e}\left( {\bf H}\times \left( \frac{\partial {\bf V}_q}{%
\partial t}\times {\bf H}\right) \right) ,
\end{equation}
where ${\bf H}$, $n_q$ are equilibrium values of the magnetic field and
quasiparticle density, while ${\bf h,}$ $n_q^{\prime }$ are small
perturbations. The last term in Eq. (\ref{m3}) appears due to the
displacement current in Maxwell's equations. Similar derivations can be done
for the hole component, they result in the same equations (\ref{m1})-(\ref
{m3}) with $n_h$, ${\bf V}_h$ instead of $n_q$, ${\bf V}_q$. Hence, one can
seek a solution describing motion of one of the components independently,
the equations of motion of the other component will be automatically
satisfied, e.g., by taking $n_h=n_q$ and ${\bf V}_h={\bf V}_q$.

Let us now derive an equation for propagation of magnetosonic waves. Taking $%
{\rm div}$ from both sides of Eq. (\ref{m3}), using Eqs. (\ref{m1}), (\ref
{m2}) and identities
$$
{\rm div}({\bf H}\times {\rm curl}{\bf h})={\rm curl}{\bf h\cdot {\rm curl}%
H-H\cdot }{\rm curl}({\rm curl}{\bf h)\approx H\cdot }\Delta {\bf h},
$$
$$
{\rm div}\left( {\bf H}\times \left( \frac{\partial {\bf V}_q}{\partial t}%
\times {\bf H}\right) \right) =-{\bf H\cdot }{\rm curl}\left( \frac{\partial
{\bf V}_q}{\partial t}\times {\bf H}\right) =-{\bf H\cdot }\frac{\partial ^2%
{\bf h}}{\partial t^2},
$$
we obtain
\begin{equation}
\label{m4}\frac{\partial ^2n_q^{\prime }}{\partial t^2}-v_F^2\Delta
n_q^{\prime }-\frac{n_q}{4\pi \rho _e}{\bf H\cdot }\Delta {\bf h}+\frac{n_q}{%
4\pi c^2\rho _e}{\bf H\cdot }\frac{\partial ^2{\bf h}}{\partial t^2}=0.
\end{equation}
The magnetic field ${\bf H}$ creates the space anisotropy in our problem. It
is convenient to decompose ${\bf h}$ and ${\bf V}_q$ into longitudinal
(parallel to ${\bf H}$) and transverse (perpendicular to ${\bf H}$)
components:%
$$
{\bf h}={\bf h}_{\parallel }+{\bf h}_{\perp },\qquad {\bf V}_q={\bf V}%
_{\parallel }+{\bf V}_{\perp }.
$$
Then from Eqs. (\ref{m1}), (\ref{m3}) we find
\begin{equation}
\label{m5}\frac{\partial h_{\parallel }}{\partial t}=-H{\rm div}{\bf V}%
_{\perp },
\end{equation}
\begin{equation}
\label{m6}\frac{\partial {\bf h}_{\perp }}{\partial t}=({\bf H\cdot }\nabla
_{\parallel }){\bf V}_{\perp },
\end{equation}
\begin{equation}
\label{m7}\frac{\partial {\bf V}_{\parallel }}{\partial t}=-\frac{v_F^2}{n_q}%
\nabla _{\parallel }n_q^{\prime }.
\end{equation}
Using Eqs. (\ref{m2}), (\ref{m5}) and (\ref{m7}), we obtain
\begin{equation}
\label{m8}\frac{\partial ^2n_q^{\prime }}{\partial t^2}-v_F^2\Delta
_{\parallel }n_q^{\prime }-\frac{n_q}H\frac{\partial ^2h_{\parallel }}{%
\partial t^2}=0,
\end{equation}
where $\Delta _{\parallel }\equiv \nabla _{\parallel }^2$. Finally, Eqs. (%
\ref{m4}), (\ref{m8}) result in the following equation for wave propagation
\begin{equation}
\label{m9}\left( 1+\frac{u_A^2}{c^2}\right) \frac{\partial ^4n_q^{\prime }}{%
\partial t^4}-\left( (v_F^2+u_A^2)\Delta +\frac{v_F^2u_A^2}{c^2}\Delta
_{\parallel }\right) \frac{\partial ^2n_q^{\prime }}{\partial t^2}%
+v_F^2u_A^2\Delta _{\parallel }\Delta n_q^{\prime }=0,
\end{equation}
where $u_A=H/\sqrt{4\pi \rho _e}$ is the Alfven velocity determined by the
total electron density.

Let us consider a plane magnetosonic wave with the wavevector ${\bf k}$: $%
n_q^{\prime }\propto \exp (i\omega t-i{\bf kr})$. Then Eq. (\ref{m9}) gives
the following expression for the speed of fast magnetosonic wave
\begin{equation}
\label{m10}u^2=\left( v_F^2+u_A^2+\frac{v_F^2u_A^2\cos ^2\vartheta }{c^2}+%
\sqrt{\left[ v_F^2+u_A^2+\frac{v_F^2u_A^2\cos ^2\vartheta }{c^2}\right]
^2-4\left( 1+\frac{u_A^2}{c^2}\right) v_F^2u_A^2\cos ^2\vartheta }\right)
/2(1+u_A^2/c^2),
\end{equation}
where $\vartheta $ is the angle between the wavevector ${\bf k}$ and the
magnetic field ${\bf H}$. Near the crust bottom $u_A\ll v_F$ and $u\approx
v_F$ (the wave reduces to zero sound) while at the star surface $u_A\gg v_F$
and $u\approx u_A/\sqrt{1+u_A^2/c^2}$ (magnetoplasma wave). We conclude that
everywhere inside the stellar crust the speed of fast magnetosonic wave is
approximately equal to the speed of light and, therefore, trajectories of
rays are approximately straight lines.

\section{Propagation of electromagnetic waves across the stellar atmosphere}

In this section we discuss conditions at which electromagnetic radiation
produced by vortices propagates across the stellar atmosphere without
attenuation. The NS atmosphere is a thin plasma layer. Due to huge
gravitational field $g\sim 10^{14}-10^{15}{\rm cm/s}^2$ at the surface, the
NS atmospheres are extremely compressed. The scale height, $\sim Zk_BT_s/Mg$
($M$ and $Z$ is the mass and charge of ions, $T_s$ is the surface
temperature), lies in the range $0.1-100${\rm cm} \cite{Pavl95}. Typical
densities of the NS atmospheres are $0.1-10{\rm g/cm}^3$. The density grows
with increasing $g$ and decreases with increasing the atmosphere
temperature. To be specific we assume ionized hydrogen atmosphere ($Z=1$, $%
m=m_e$, $M=m_p$). A strong magnetic field of the order of $10^{12}-10^{13}%
{\rm Gs}$ can exist near the star magnetic poles. In such magnetic field the
cyclotron frequency of electrons is (we take in numerical estimates $%
H=4\times 10^{12}{\rm Gs}$)
\begin{equation}
\label{p1}\omega _{He}=\frac{|e|H}{mc}=1.1\times 10^{19}{\rm s}^{-1},
\end{equation}
while the ion cyclotron frequency
\begin{equation}
\label{p2}\omega _{Hi}=\frac{|e|H}{Mc}=6\times 10^{15}{\rm s}^{-1},
\end{equation}
here $e$ is the electron charge. The strong magnetic field drastically
distort the structure of atoms and increases the binding energies of
electrons in atoms. For instance, the field $10^{12}{\rm Gs}$ amplifies the
binding energy of hydrogen atom to $\sim 150{\rm eV}\approx 1.7\times 10^6$%
K. For lower temperatures the magnetized atmospheres become partly ionized.
In this section we assume that the atmosphere temperature is greater than $%
10^6$K so that the magnetoactive plasma is highly ionized with $n_e=n_i=n$ ($%
n_e$ and $n_i$ are electron and ion concentrations). In numerical estimates
we take $T_s=2\times 10^6$K.

An effective collision frequency of particles characterizes the plasma. For
parameters of the stellar atmosphere $\omega _{He}\gg k_BT_s/\hbar \approx
4\times 10^{16}$Hz that is thermal energy is much less then the spacing
between Landau levels of electron in magnetic field. Therefore, collisions
of an electron with other particles are substantially suppressed because the
electron motion is effectively one dimensional (only momentum along the
magnetic field can be changed during the collision). Contrary to electrons,
spacing between Landau levels of ions is small in comparison with the
temperature ($\hbar \omega _{Hi}\ll k_BT_s$) and, as a result, the ion-ion
collisions are dominant. The effective ion-ion collision frequency can be
estimated as \cite{Ginz70}
\begin{equation}
\label{p3}\nu _{{\rm eff}}\approx \frac{\pi e^4}{\sqrt{2}(k_BT_s)^2}\bar v%
_in_i\ln \left( \frac{k_BT_s}{e^2n_i^{1/3}}\right) =\frac{3.9n_i}{T_s^{3/2}}%
\sqrt{\frac mM}\ln \left( \frac{220T_s}{n_i^{1/3}}\right) ,
\end{equation}
here $\bar v_i$ is the average ion velocity. Magnetic field does not affect
the nature of collisions provided $r_H\gg r_D$, where $r_H=\bar v_i/\omega
_{Hi}$ is the radius of curvature of a particle in the field and $r_D\sim
\bar v_i/\omega _{pi}$ is the Debye radius. From this we arrive at condition
$\omega _{pi}\gg \omega _{Hi}$. If this condition is not satisfied (as in
our problem), this fact, however, affects only the logarithmic factor in (%
\ref{p3}) and Eq. (\ref{p3}) still can be used for approximate estimation of
the collision frequency. For $n_i=10^{24}{\rm cm}^{-3}$, $T_s=2\times 10^6$K
we obtain that $\nu _{{\rm eff}}$ lies in the infrared band: $\nu _{{\rm eff}%
}\approx 4.7\times 10^{13}{\rm Hz}$. For colder NSs the atmosphere
temperature is less than $\sim 10^6$K, plasma is weakly ionized and ion-atom
collisions become dominant.

For $n=10^{24}{\rm cm}^{-3}$ the plasma frequencies are%
$$
\omega _{pe}=\sqrt{\frac{4\pi e^2n}m}=9\times 10^{15}{\rm s}^{-1},
$$
$$
\omega _{pi}=\sqrt{\frac{4\pi e^2n}M}=2.1\times 10^{14}{\rm s}^{-1}.
$$

We consider electromagnetic waves in plasma under the condition $\omega \gg
\nu _{{\rm eff}}$ (collisionless plasma) which suggests their small damping.
The opposite limit $\omega \ll \nu _{{\rm eff}}$ corresponds to magnetic
hydrodynamics. We choose a coordinate system so that the $z$ axis is
oriented along the magnetic field ${\bf H}$. The dielectric permittivity
tensor of a cold collisionless plasma in a magnetic field has the form \cite
{Lifs79}

\begin{equation}
\label{f1}\varepsilon _{xx}=\varepsilon _{yy}=\varepsilon _{\perp },\quad
\varepsilon _{zz}=\varepsilon _{\parallel },
\end{equation}
\begin{equation}
\label{f2}\varepsilon _{xy}=-\varepsilon _{yx}=i\eta ,\quad \varepsilon
_{xz}=\varepsilon _{yz}=0,
\end{equation}
\begin{equation}
\label{f3}\varepsilon _{\perp }=1-\frac{\omega _{pe}^2}{\omega ^2-\omega
_{He}^2}-\frac{\omega _{pi}^2}{\omega ^2-\omega _{Hi}^2},
\end{equation}
\begin{equation}
\label{f4}\varepsilon _{\parallel }=1-\frac{\omega _{pe}^2+\omega _{pi}^2}{%
\omega ^2},
\end{equation}
\begin{equation}
\label{f5}\eta =\frac{\omega _{He}\omega _{pe}^2}{\omega (\omega ^2-\omega
_{He}^2)}-\frac{\omega _{Hi}\omega _{pi}^2}{\omega (\omega ^2-\omega _{Hi}^2)%
}.
\end{equation}
The dispersion equation for electromagnetic waves
\begin{equation}
\label{f9}\left| k^2\delta _{\alpha \beta }-k_\alpha k_\beta -\frac{\omega ^2%
}{c^2}\varepsilon _{\alpha \beta }\right| =0
\end{equation}
has two solutions. One of them describes fast magnetosonic waves and for
frequencies not too close to cyclotron frequencies $\omega _{Hi}$, $\omega
_{He}$ has the form
\begin{equation}
\label{f10}\omega =\frac{kc}{\sqrt{\varepsilon _{\perp }}}.
\end{equation}
The other solution corresponds to Alfven branch. General expression for the
dispersion relation of this branch reduces to a simple form only in limiting
cases. E. g., in the frequency range $\nu _{{\rm eff}}\ll \omega \ll \omega
_{Hi},\omega _{pe}$ the dispersion equation for Alfven waves is
\begin{equation}
\label{f11}\omega =\frac{c\omega _{pe}k\cos \vartheta }{\sqrt{\varepsilon
_{\perp }}\sqrt{(kc\sin \vartheta )^2+\omega _{pe}^2}},
\end{equation}
where $\vartheta $ is the angle between ${\bf k}$ and ${\bf H}$.

Further we consider the fast magnetosonic branch only because this type of
waves is generated by vortices. Ion-ion collisions produce damping of
electromagnetic waves. To estimate parameters at which the atmosphere is
transparent for fast magnetosonic waves one can use the following expression
for the complex dielectric permittivity \cite{Ginz70}
\begin{equation}
\label{f12}\varepsilon _{\perp }=1-\frac{\omega _{pe}^2}{\omega ^2-\omega
_{He}^2}-\frac{\omega _{pi}^2(\omega +i\nu _{{\rm eff}})}{\omega \left[
(\omega +i\nu _{{\rm eff}})^2-\omega _{Hi}^2\right] }\approx 1-\frac{\omega
_{pe}^2}{\omega ^2-\omega _{He}^2}-\frac{\omega _{pi}^2}{\omega ^2-\omega
_{Hi}^2}+\frac{i\nu _{{\rm eff}}}\omega \frac{\omega _{pi}^2(\omega
^2+\omega _{Hi}^2)}{(\omega ^2-\omega _{Hi}^2)^2}.
\end{equation}
At frequencies not too close to cyclotron frequencies the real part of the
refractive index is ${\rm Re}t\approx 1$, while the imaginary part is
\begin{equation}
\label{f13}{\rm Im}t={\rm Im}\sqrt{\varepsilon _{\perp }}\approx \frac{\nu _{%
{\rm eff}}}{2\omega }\frac{\omega _{pi}^2(\omega ^2+\omega _{Hi}^2)}{(\omega
^2-\omega _{Hi}^2)^2}.
\end{equation}
If radiation propagates through the atmosphere along the $z$ axis its
amplitude decreases by a factor $\exp (-\omega \int {\rm Im}tdz/c)$. In
estimates we assume an isothermal atmosphere. Deviation of the atmosphere
from an isothermal one does not change the result substantially. Then
concentration of particles changes with the height $z$ as $n(z)=n_0\exp
(-Mgz/k_BT_s)$, where $n_0$ is the concentration at the atmosphere bottom ($%
z=0$). The effective collision frequency $\nu _{{\rm eff}}\propto n(z)$,
and, therefore, ${\rm Im}t\propto n^2$ also decreases with the height
according to the exponential dependence. Using the criteria $\omega \int
{\rm Im}tdz/c\approx 1$ one can estimate the maximum density at the
atmosphere bottom for which the atmosphere is still transparent for vortex
radiation
\begin{equation}
\label{p4}n_0\nu _{{\rm eff}}\approx \frac{cgM^2}{\pi e^2k_BT_s}\frac{%
(\omega ^2-\omega _{Hi}^2)^2}{(\omega ^2+\omega _{Hi}^2)}.
\end{equation}
Comparing Eqs. (\ref{p4}) and (\ref{p3}), we finally obtain
\begin{equation}
\label{p5}n_0\approx \frac{M^{1/4}g^{1/2}T_s^{1/4}H}{m^{1/4}\sqrt{3.9\pi \ln
(220T_s/n^{1/3})ck_B}}\left\{
\begin{array}{c}
1,\quad \omega \ll \omega _{Hi} \\
\omega /\omega _{Hi},\quad \omega \gg \omega _{Hi}
\end{array}
\right. .
\end{equation}
In the limit $\omega \ll \omega _{Hi}$ the maximum value of $n_0$ is
independent of $\omega $. If $g=10^{15}{\rm cm/s}^2$, $T_s=2\times 10^6$K
and $H=4\times 10^{12}{\rm Gs}$ we obtain $n_0\approx 4.4\times 10^{24}{\rm %
cm}^{-3}$. In this limit Eq. (\ref{p5}) approximately coincides with the
photosphere density obtained from numerical calculations assuming free-free
opacity \cite{Lai92}: $n_{{\rm phot}}(H)\approx 1.8\cdot
10^{23}g_{14}^{1/2}T_5^{1/4}H_{12}=4.8\times 10^{24}{\rm cm}^{-3}$. For $%
\omega >\omega _{Hi}$ the maximum value of $n_0$ becomes frequency dependent
and increases with increasing $\omega $. We conclude that if the particle
concentration at the atmosphere bottom is less than $4\times 10^{24}{\rm cm}%
^{-3}$ (which is usually the case) the atmosphere is transparent for vortex
radiation (apart from narrow regions near cyclotron frequencies).

\begin{figure}
\bigskip
\centerline{\epsfxsize=0.37\textwidth\epsfysize=0.37\textwidth
\epsfbox{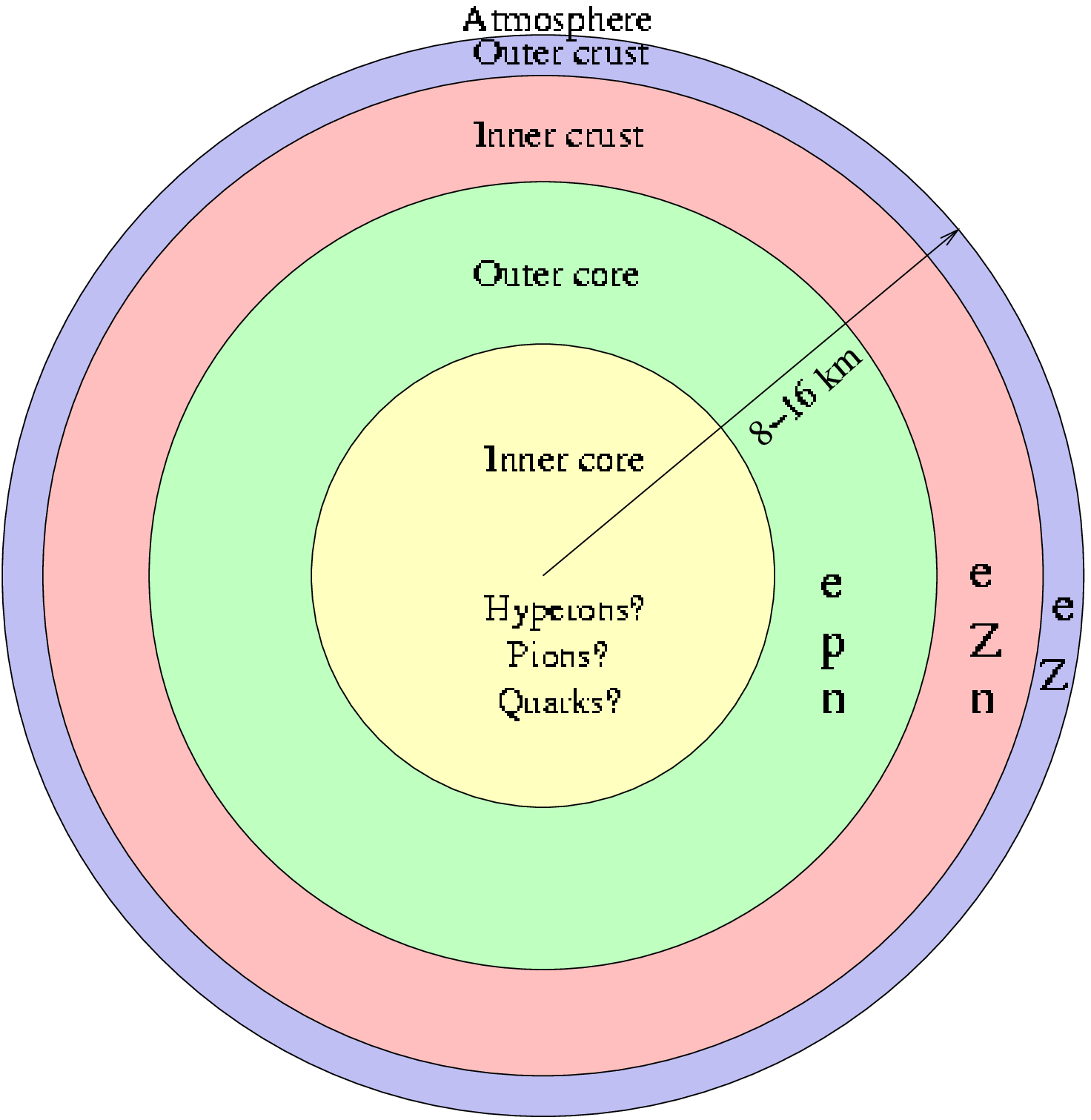}}

\caption{Schematic cross section of a dense neutron star. The star
is subdivided into the atmosphere, the outer crust, the inner
crust and superfluid outer and inner cores. }
\label{fig1}
\end{figure}

\begin{figure}
\bigskip
\centerline{\epsfxsize=0.35\textwidth\epsfysize=0.43\textwidth
\epsfbox{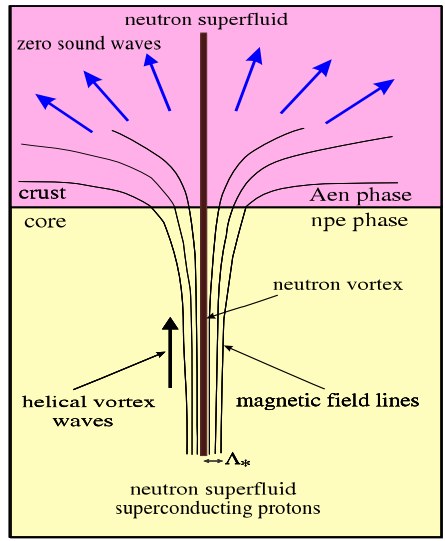}}

\caption{Generation of zero sound waves by vortices at the
crust-core interface. Due to drag effect the helical vortex motion causes
density oscillation of electrons in the stellar core. No sound is radiated
inside the core because the wave length of sound is larger then the helix
period. However, at the crust-core interface the density oscillation excites
spherical sound waves that propagate across the stellar crust.
}
\label{fig2}
\end{figure}

\begin{figure}
\bigskip
\centerline{\epsfxsize=0.45\textwidth\epsfysize=0.40\textwidth
\epsfbox{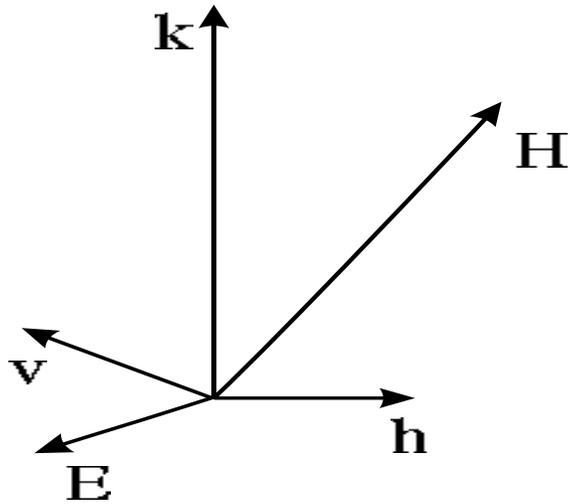}}

\vspace{0.6cm}

\caption{Oscillations of electric and magnetic fields in the fast magnetosonic
wave in the limit $u_A \gg u_s$. Oscillating magnetic ${\bf h}$ and
electric fields ${\bf E}$ are perpendicular to each other and the
wavevector ${\bf k}$. The electron velocity ${\bf V}$ lies in the
$kH$ plane and perpendicular to ${\bf H}$.
}
\label{fig3}
\end{figure}

\begin{figure}
\bigskip
\centerline{\epsfxsize=0.55\textwidth\epsfysize=0.75\textwidth
\epsfbox{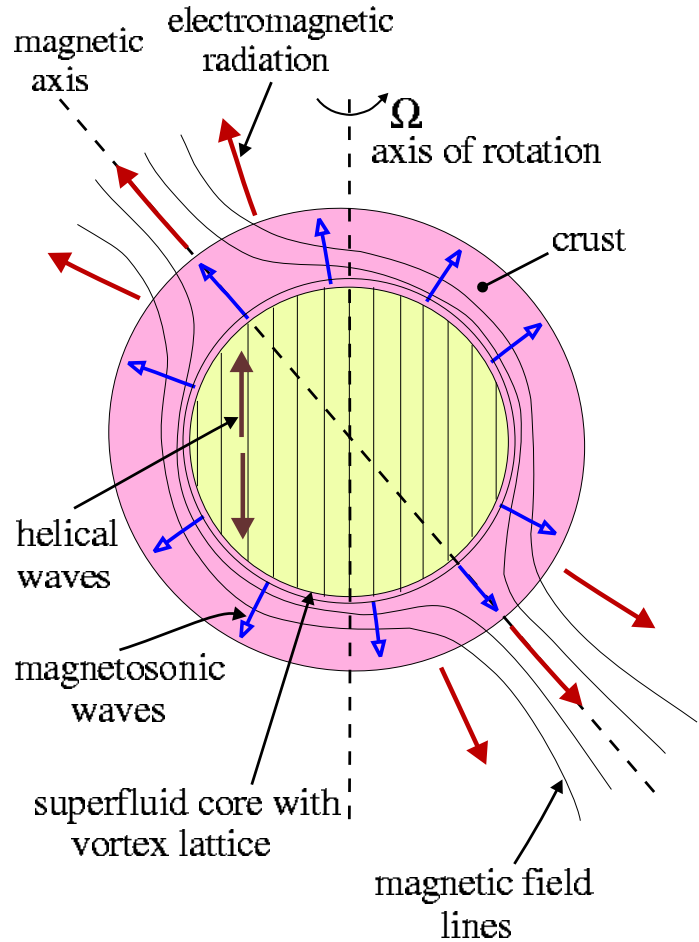}}

\vspace{0.6cm}

\caption{Mechanism of a neutron star radiation. Thermaly excited helical waves
of neutron vortices in the superfluid core produce magnetosonic waves in the
stellar crust. Magnetosonic waves propagate across the crust and transform into
electromagnetic radiation at the star surface. Mainly the radiation comes out
from regions with strong magnetic field (near magnetic poles).}
\label{fig4}
\end{figure}

\begin{figure}
\bigskip
\centerline{\epsfxsize=0.5\textwidth\epsfysize=0.60\textwidth
\epsfbox{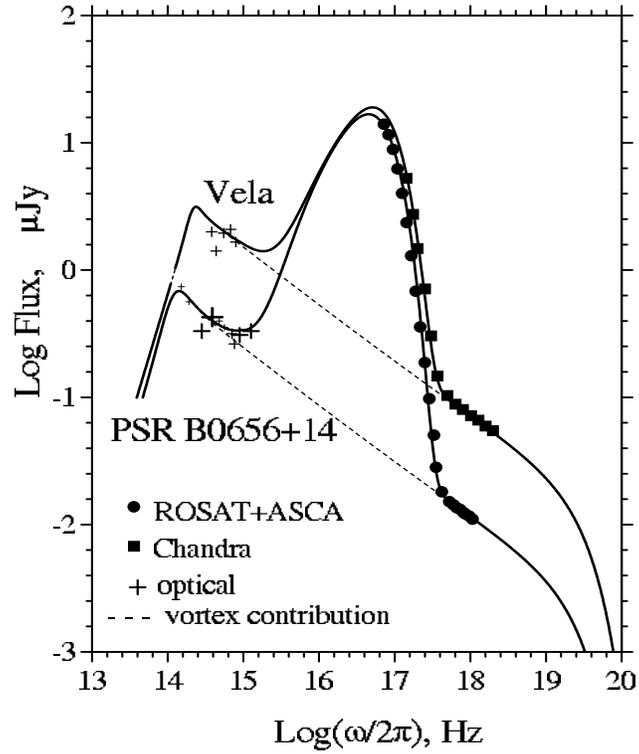}}

\vspace{0.6cm}

\caption{Broadband spectrum of PSR B0656+14 (Koptsevich {\rm et~al.} 2000) and
the Vela pulsar (Pavlov {\rm et~al.} 2001, Nasuti {\rm et~al.} 1997). Solid line
is
the fit by the sum of the vortex and thermal components. Thermal radiation of
the stellar surface dominates in the ultraviolet and soft $X-$ray bands, while
the vortex contribution (dash line) prevails in infrared, optical and hard
$X-$ray bands, where its spectrum has a slope $\alpha \approx -0.45$.
In the far infrared band the spectrum of vortex radiation changes its
behavior and follows Planck's formula with $P(\omega)\propto \omega^2$.
}
\label{fig5}
\end{figure}

\begin{figure}
\bigskip
\centerline{\epsfxsize=0.5\textwidth\epsfysize=0.60\textwidth
\epsfbox{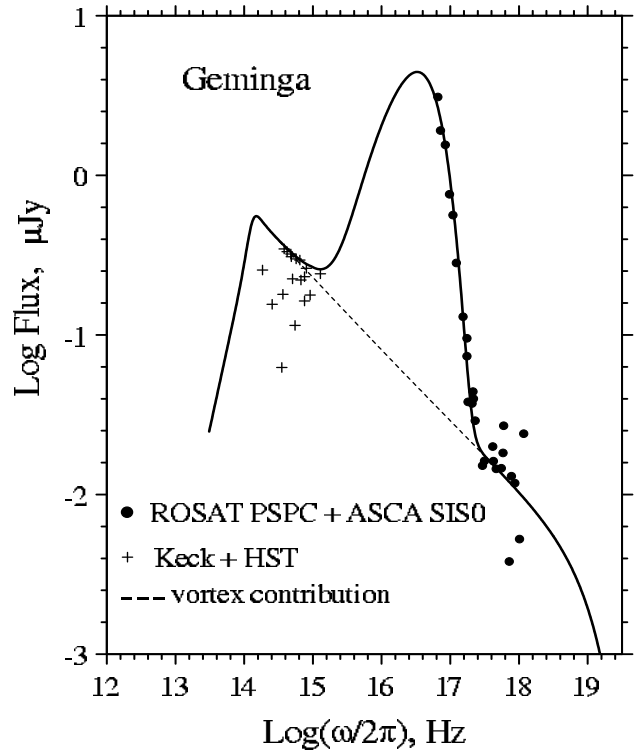}}

\vspace{0.6cm}

\caption{Broadband spectrum of the Geminga pulsar.
Filled circles are data of ROSAT and ASCA (Halpern \& Wang 1997), while crosses
combine measurements of the Keck Observatory (Martin {\rm et~al.} 1998) and
the Hubble Space Telescope (Mignani {\rm et~al.} 1998). Solid line is the fit
by the sum of the vortex (dash line) and thermal contributions.}
\label{fig6}
\end{figure}

\begin{figure}
\bigskip
\centerline{\epsfxsize=0.45\textwidth\epsfysize=0.50\textwidth
\epsfbox{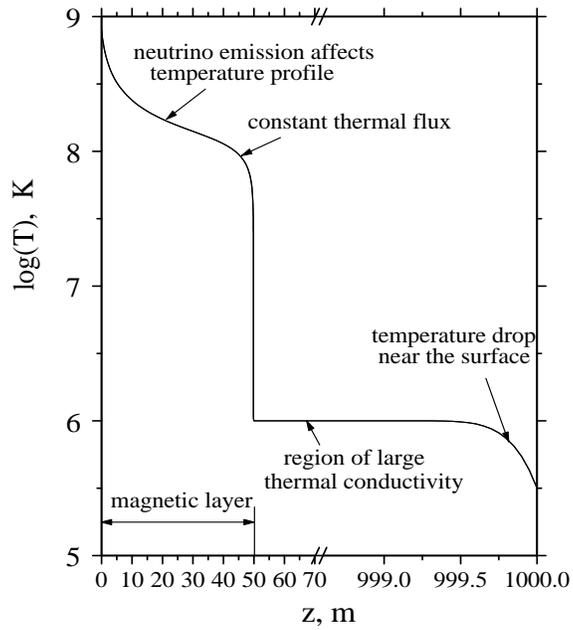}}

\vspace{0.6cm}

\caption{Temperature distribution in a neutron star crust for a model with
magnetic isolating layer of width $d=50$m. The main temperature
drop occurs near the crust bottom where temperature decreases from
$T=8\times 10^8$K (at the core-crust interface, $z=0$) to $10^6$K. The rest
part of the crust is approximately isothermal apart from a thin surface layer
where temperature drops due to substantial decrease of thermoconductivity
at low densities. Near the crust bottom the neutrino emissin results in energy
losses and only a small fraction of the original heat flux
reaches the stellar surface.}
\label{fig7}
\end{figure}

\end{document}